\documentclass[fleqn,usenatbib]{mnras}
\usepackage{float}
\usepackage{newtxtext,newtxmath}
\usepackage{afterpage}
\usepackage{placeins}
\usepackage[T1]{fontenc}
\usepackage{xcolor}
\DeclareRobustCommand{\VAN}[3]{#2}
\let\VANthebibliography\thebibliography
\def\thebibliography{\DeclareRobustCommand{\VAN}[3]{##3}\VANthebibliography}

\usepackage{orcidlink}

\usepackage{graphicx}	
\usepackage{amsmath}	

\title[ The strong \textasciitilde2.5 Hz modulation in MAXI J1816--195]{Detection of a strong \textasciitilde2.5 Hz modulation in the Newly Discovered Millisecond Pulsar MAXI J1816--195}

\author[Panping Li et al.]{
P. P. Li,$^{1,2}$ L. Tao,$^{1}$\thanks{E-mail: taolian@ihep.ac.cn} L. Zhang,$^{1}$ Q. C. Bu~\orcidlink{0000-0001-5238-3988},$^{3}$ J. L. Qu,$^{1}$ L. Ji~\orcidlink{0000-0001-9599-7285},$^{4}$ P. J. Wang,$^{1,2}$ Y. P. Chen~\orcidlink{0000-0001-8768-3294},$^{1}$  S. Zhang,$^{1}$
\newauthor{R. C. Ma,$^{1, 2}$ Z. X. Yang,$^{1, 2}$ W. T. Ye,$^{1, 2}$ S. J. Zhao,$^{1, 2}$ Q. C. Zhao,$^{1, 2}$ Y. Huang,$^{1}$ X. Ma,$^{1}$  E. L. Qiao,$^{5}$ } 
\newauthor{S. M. Jia,$^{1}$ S. N. Zhang$^{1}$}
\\
$^{1}$Key Laboratory of Particle Astrophysics, Institute of High Energy Physics, Chinese Academy of Sciences, 100049 Beijing, China\\
$^{2}$Uinversity of Chinese Academy of Sciences, Chinese Academy of Sciences, 100049 Beijing, China\\
$^{3}$Institut f\"ur Astronomie und Astrophysik, Kepler Center for Astro and Particle Physics, Eberhard Karls Universit\"at, Sand 1, D-72076 T\"ubingen, Germany\\
$^{4}$School of Physics and Astronomy, Sun Yat-sen University, Zhuhai, 519082, People’s Republic of China\\
$^{5}$National Astronomical Observatories, Chinese Academy of Sciences, Beijing 100101, China}

\date{Accepted XXX. Received YYY; in original form ZZZ}

\pubyear{2015}

\begin{document}
\label{firstpage}
\pagerange{\pageref{firstpage}--\pageref{lastpage}}
\maketitle

\begin{abstract}
MAXI J1816--195 is a newly discovered accreting millisecond X-ray pulsar that went outburst in June 2022. Through timing analysis with \textit{NICER} and \textit{NuSTAR} observations, we find a transient modulation at \textasciitilde2.5 Hz during the decay period of MAXI J1816--195. The modulation is strongly correlated with a spectral hardening, and its fractional rms amplitude increases with energy. These results suggest that the modulation is likely to be produced in an unstable corona. In addition, the presence of the modulation during thermonuclear bursts indicates that it may originate from a disk-corona where the optical depth is likely the main factor affecting the modulation, rather than temperature. Moreover, we find significant reflection features in the spectra observed simultaneously by \textit{NICER} and \textit{NuSTAR}, including a relativistically broadened Fe-K line around 6--7\,keV, and a Compton hump in the 10--30\,keV energy band. The radius of the inner disc is constrained to be $R_{\rm in}$ = (1.04--1.23)\,$R_{\rm ISCO}$ based on reflection modeling of the broadband spectra. Assuming that the inner disc is truncated at the magnetosphere radius, we estimate that the magnetic field strength is $\leq 4.67 \times 10^{8}\,\rm G$. 
\end{abstract}




\section{Introduction}

 Low-mass X-ray binary systems (LMXBs) consist of a compact object, which is either a black hole (BH) or a neutron star (NS), and a companion star (mass $< 1\ M_{\odot}$) which emits X-ray radiation resulting from the release of gravitational potential energy from the fall of accreting material via Roche-lobe \citep[]{1975van}. A NS-LMXB is usually accompanied by an old neutron star with a weak magnetic field. Thus, the behavior of NS-LMXBs is similar to that of BH-LMXBs in terms of variability and state transition. However, the presence of rigid surfaces and natural magnetic fields in NS-LMXBs, as well as smaller gravitational forces, distinguish them from BH-LMXBs (see \cite{2013Matsuoka} and references therein).

Accreting millisecond X-ray pulsars (AMXPs) are a special subclass of NS-LMXBs with spin frequencies of several hundred hertz (see reviews by \cite{2020Di}). Since the discovery of the first accretion millisecond pulsar, SAX J1808.4--3658, by \textit{RXTE} in 1998 \citep[]{1998Wijnands}, a total of 24 accretion millisecond pulsars have been discovered \citep[]{2022Marino}, with an average of one discovery per year. Most recently, two accretion millisecond pulsars, MAXI J1816--195 \citep[]{2022Negoro} and MAXI J1957+032 \citep[]{2022ATelNg}, were discovered in June 2022. All AMXPs are transients, usually with X-ray outbursts recurring every 2--4 yrs. The outburst duration is typically a couple of weeks, reaching luminosities of $10^{36 - 37}\ \rm ergs\ \rm s^{-1}$ in the X-ray band \citep[]{2020Kuiper}. As with some other NS-LMXBs, the spectra of AMXPs in outbursts generally include a thermal component emitted by an accretion disc, the NS surface, or a boundary layer, and a high-energy nonthermal component produced by the hot corona of tens of keV via the scattering of soft photons. The majority of AMXPs also contain a reflection component of the hard Comptonized photons illuminating the cold accretion disk, consisting of several emission lines and a broad hump of 10--30\,keV \citep{2013Papitto,2016Pintore,2022Marino}. The Fe-K line is the most prominent feature in the reflection spectrum, which in turn provides constraints on the structure of the inner disc and inclination \citep[]{1989Fabian}.

In addition, the X-ray emission from AMXPs also shows rich variability. These various timing features in the power-density spectra (PDS) of AMXPs are commonly present as broad-band noise continuum and sharp peaks called quasi-periodic oscillations (QPOs). The study of QPOs or similar modulations reveals a rich phenomenology which has helped us to understand the accretion physics of compact objects \citep[e.g.][]{2021Belloni}. The kHz QPO has been found in many millisecond pulsars, such as SAX J1808.4--3658 \citep[]{2003Wijnands}, XTE J1807--294 \citep[]{2005Linares}, IGR J17511--305 \citep[]{2011Kalamkar} and Aql X--1 \citep[]{2008Barret}. Sometimes two kHz QPOs are observed simultaneously and the frequency difference between the two QPOs ($\Delta v$) is related to the NS spin frequency $v_{\rm spin}$, i.e. $\Delta v=v_{\rm spin}$ \citep[]{2005Linares} or $\Delta v=v_{\rm spin}/2$ \citep[]{2011Kalamkar}, which is consistent with the predictions of the relativistic precession model \cite{1999Stella} and  spin-resonance model \cite{2003Lamb}.


Compared with kHz QPOs, the detection of QPOs or modulations at low frequencies in AMXPs is relatively scarce. The 1\,Hz modulations of SAX J1808.4--3658 \citep{patruno2009} and NGC 6440 X--2 \citep{patruno2013} are widely known, which are interpreted as variability being caused by disk availability when the inner edge of the accretion disk is close to the co-rotation radius \citep{patruno2013}. In addition, the mHz QPOs are found in several sources, such as the \textasciitilde 6--7\,mHz QPO in Aql X--1  \citep[]{2001Revnivtsev} and the \textasciitilde 8\,mHz QPO in IGR J00291+5934 \citep[]{2017Ferrigno}. The power spectra of these sources are more complicated, especially at low frequencies. According to the characteristic frequencies of different components in the power spectra and the correlations between these characteristic frequencies, these components are generally identified as break, low-frequency QPO, harmonic of the low-frequency QPO, hump, hectoHz QPO, lower kHz QPO, or upper kHz QPO \citep[]{2005Straaten}. Moreover, similar to atoll sources, the correlations between the characteristic frequencies for AMXPs usually follow the WK \citep[]{1999Wijnands} and PBK \citep[]{1999Psaltis} relations, but being scaled by a constant factor when comparing with the atoll sources \citep[]{2005Linares,2015Bult,2017Doesburgh}. 


The low-frequency behaviors, especially QPOs or modulations, are important probes to understand the accretion variations in the outer regions and large timescales. Their studies will provide a comprehensive understanding of the physical processes of accretion in AMXPs, on the basis of high-frequency behaviors. However, as mentioned above, the low-frequency QPOs or modulations are only observed in a few AMXPs, thus it is very important to obtain new observational samples and perform in-depth studies. 

MAXI J1816--195 is a new accreting millisecond X-ray pulsar discovered by \textit{MAXI}/GSC on June 07, 2022 \citep[]{2022Negoro}. Some X-ray satellites perform follow-up observations. With \textit{NICER} observations, a pulsation of 528\,Hz \citep[]{2022BultA} and an orbital period of 17402.27\,s are found \citep[]{2022BultB}. Moreover, \textit{NuSTAR} observed this source on June 23, 2022 with a total exposure time of 40\,ks \citep[]{2022Chauhan}. By analyzing this \textit{NuSTAR} observation and the simultaneous \textit{NICER} observation, \cite{2022Mandal} studied the evolution of the type-I burst profile with flux and energy. \textit{Insight}-HXMT also observed this source during the outburst \citep[]{2022Li}, and interestingly, 73 thermonuclear X-ray bursts have been detected. \cite{2022Chen} selected 66 bursts with similar burst profiles and intensities, and found significant hard X-ray shortages during the bursts due to a burst cooling effect. Moreover, \cite{2023Li} reported the pulse profile and the spin evolution using the \textit{Insight}-HXMT observations, revealing the detection of X-ray pulsations with energies up to $\sim$95\,keV, and found that the pulse profiles were characterized by a truncated Fourier series with two harmonics.



In this paper, we will study the timing and spectral properties of MAXI J1816--195 using data from \textit{NuSTAR} and \textit{NICER} observations, and focus on the behaviors of the \textasciitilde2.5\,Hz modulation. The paper is structured as follows. Observation and data reduction are presented in Section~\ref{sec:2}, and the results are summarised in Section~\ref{sec:3}. The discussion and conclusions are presented in Section~\ref{sec:4}.


\begin{figure*}
	\includegraphics[width=0.38\textwidth]{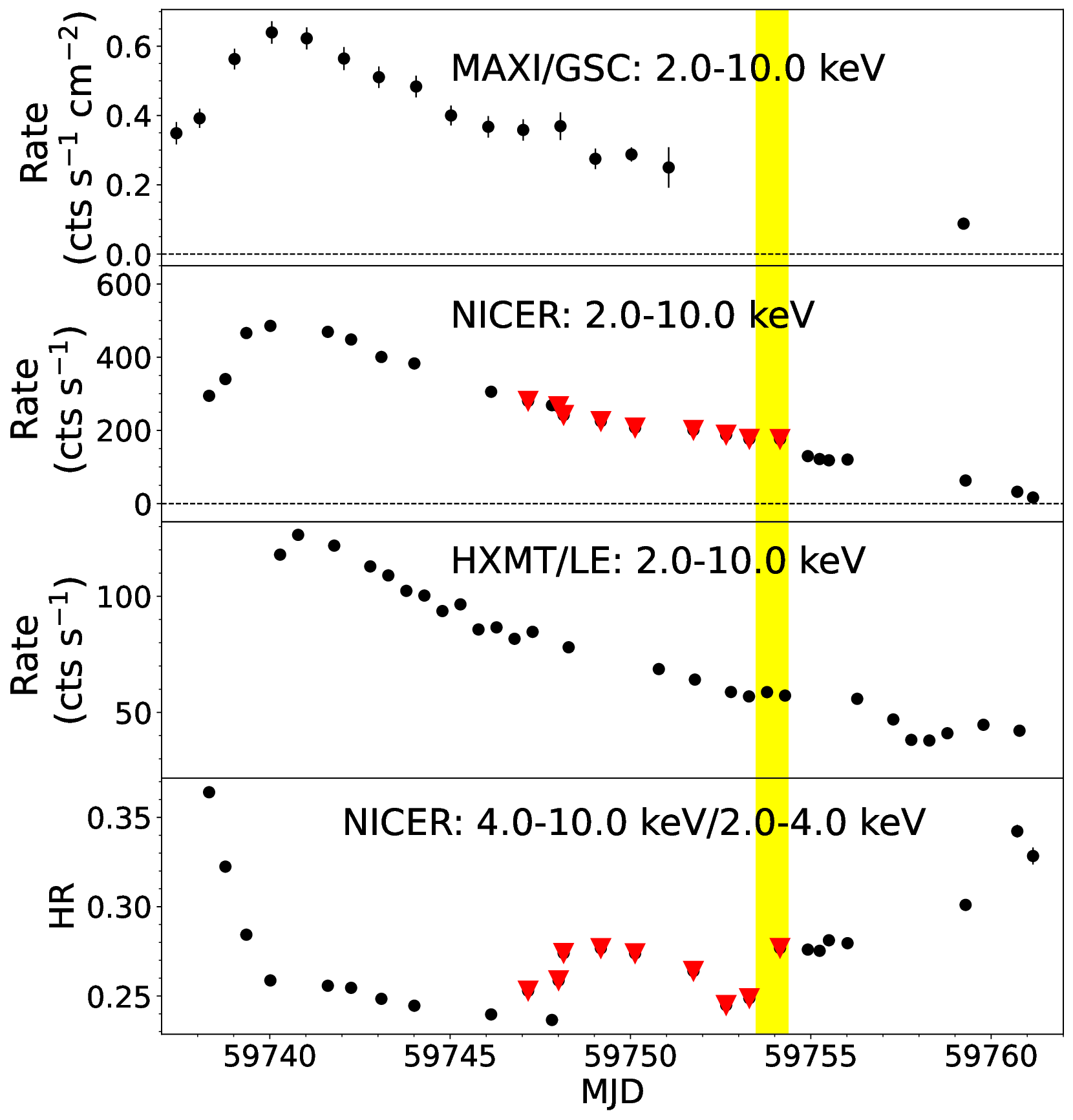}
	\includegraphics[width=0.61\textwidth]{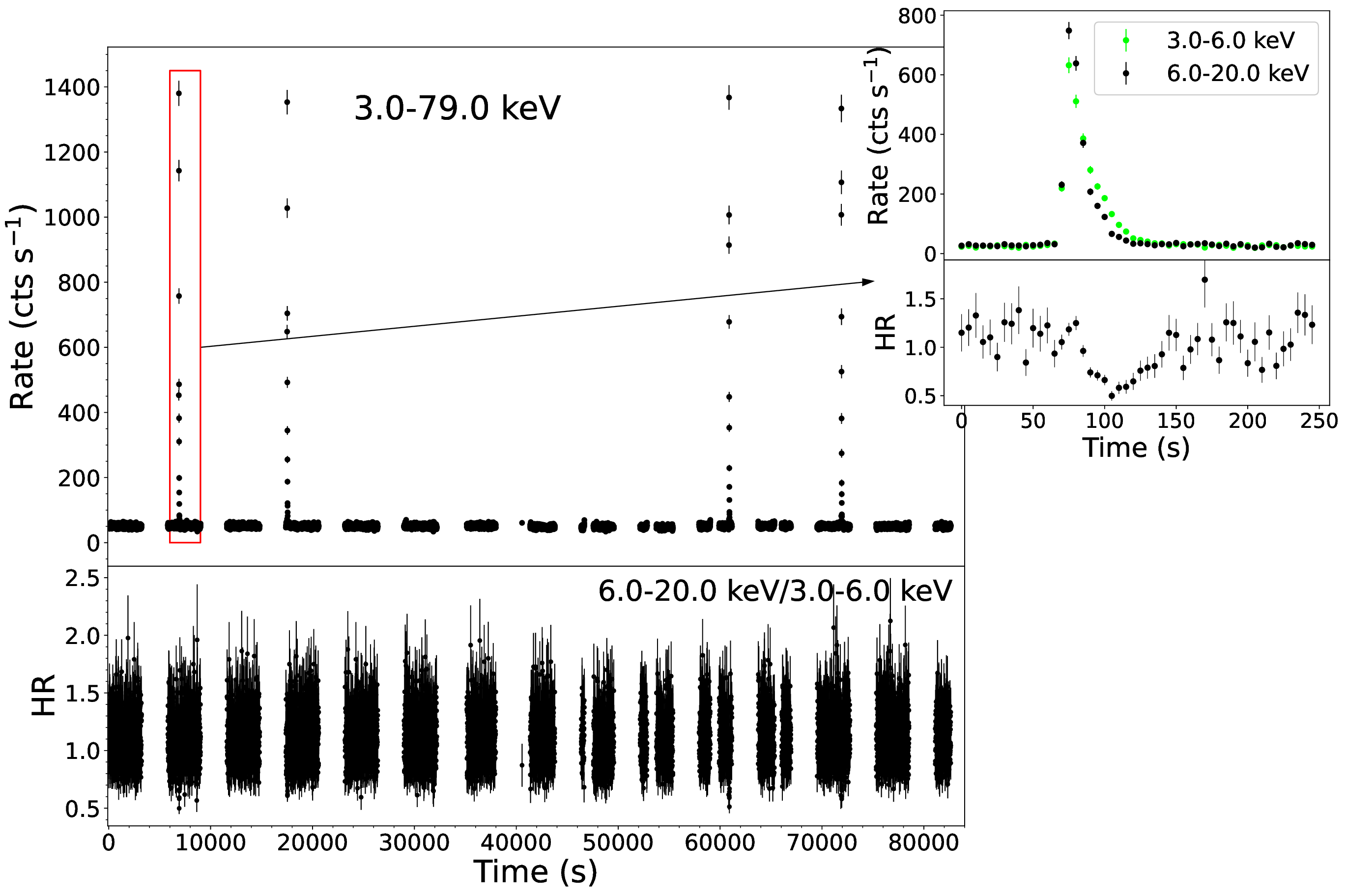}
    \caption{X-ray light curves and hardness ratios of MAXI J1816--195 during the 2022 outburst. Left panels from top to bottom: the 2--10\,keV light curves of \textit{MAXI}/GSC (1-day bin), \textit{NICER} (1-ObsID bin) and \textit{Insight}-HXMT/LE (0.5-day bin), and the hardness ratio between the 4--10 keV and 2--4 keV bands using the \textit{NICER} data. The \textit{NICER} observations with the \textasciitilde2.5\,Hz modulation detected are marked with red triangles. The yellow shaded area marks the time interval of \textit{NuSTAR} observations. Right panels: the 3--79 keV \textit{NuSTAR}/FPMA light curve and hardness ratio (6--20 keV/3--6 keV) with a binsize of 5\,s. The insert zooms in on the light curve and hardness ratio of the first thermonuclear burst detected by \textit{NuSTAR}.}
  \label{fig:hr_lc_1}
\end{figure*}

\section{Observations and Data Reduction}
\label{sec:2}

\subsection{\textit{NICER}}
\label{sec:2.1} 

The Neutron star Interior Composition Explorer (\textit{NICER}) is an instrument onboard the International Space Station (ISS) designed to study neutron stars in the 0.2--12 keV X-ray band \citep[]{2016SPIE}. \textit{NICER}’s public observations of MAXI J1816–195 from June 8 (ObsID 5202820102) to July 1 (ObsID 5533012301), 2022, are used in this work (Table~\ref{tab:obsid_1}). These data are processed with {\tt HEASoft v 6.30.1} and the \textit{NICER} Data Analysis Software ({\tt NICERDAS v10}) with Calibration Database ({\tt CALDB vXTI20210707}). All standard calibration and screening criteria are applied to produce clean event files from the {\tt nicerl2} tool. We use the task {\tt barycorr} to apply barycentric corrections for each clean event file. Based on the timing windows of 100 s before and after the burst peak, the event files during the burst and the non-burst (i.e. persistent emission) are extracted separately.

We choose to analyze the spectrum of ObsID 5533011601, which is synchronized with \textit{NuSTAR} (ObsID 90801315001). The total persistent spectra and background spectra are generated by the tool {\tt nibackgen3C50} \citep[]{2022Remillard}. The ancillary response files (ARFs; detectors 14 and 34 are removed) and response matrix files (RMFs) are extracted respectively by using the {\tt nicerarf} and {\tt nicermf}. Next, we use the {\tt ftgrouppha} \citep[]{2016Kaastra} command to rebin the spectrum with groupscale to 30. Spectral data are analyzed using {\tt XSPEC v12.12.1}, the system error is set to 1\%, and the energy range is limited to 0.5--10.0 keV, which was recommended by \textit{NICER} Calibration Recommendations \footnote{\url{https://heasarc.gsfc.nasa.gov/docs/nicer/analysis_threads/cal-recommend/}}.




\subsection{\textit{NuSTAR}}
\label{sec:2.2}

Between June 23 and 24, 2022, the Nuclear Spectroscopic Telescope Array (\textit{NuSTAR}) observed MAXI J1816--195 (ObsID 90801315001) with a total exposure of 40\,ks. \textit{NuSTAR}’s focal plane consists of two independent solid-state detectors (FPMA and FPMB) operating in 3--79\,kev \citep[]{2013Harrison}. The data from the \textit{NuSTAR} detectors (FPMA and FPMB) are calibrated and screened to get clean events using the {\tt nupipeline} routine of \textit{NuSTAR} data analysis software ({\tt NustarDAS v2.1.2}) with {\tt CALDB v20220706}, which is distributed with {\tt HEASoft v6.30.1}. Since the burst peak count rate exceeds 100 counts s$^{-1}$, the keyword STATUEXPR is set to be "STATUS==b0000xxx00xx0000". We use {\tt DS9} to produce separately the source (a $210^{\prime \prime}$ circular region) and background (a $110^{\prime \prime}$ circular region) extraction region files, and then apply the {\tt nuproducts} command for extracting the source spectra, light-curve, and instrument responses. The barycenter corrected cleaned event file is got by setting the parameter "write\_baryevtfile=yes" and "barycorr=yes". We also alter GTI files to separate the burst and persistent X-ray emission from MAXI J1816--195. The persistent emission is rebinned with a minimum of 50 counts per energy bin and limited to the 3--40\,keV energy range for the poor signal-to-noise ratio at high energies.

\begin{figure*}
	\includegraphics[width=\textwidth]{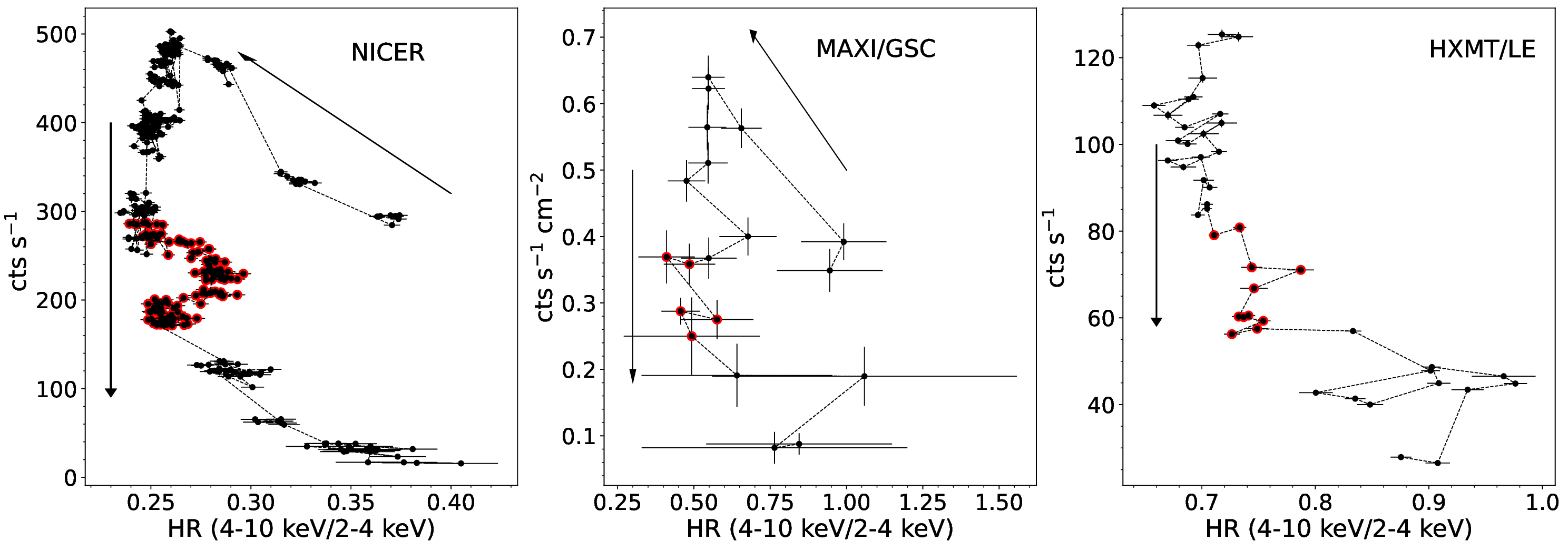}
    \caption{Left panel: The hardness-intensity density (HID) based on the data from \textit{NICER} (150\,s bins). The black arrows describe the direction of evolution. Observations using \textit{NICER} with the \textasciitilde2.5\,Hz modulation are indicated by red points. Middle panel: HID from \textit{MAXI}/GSC (1-d bins). Right panel: HID from \textit{Insight}-HXMT/LE (1-Exposure bins). The y-axes of all panels are based on photons in the 2--10 keV energy range. The dates of \textit{MAXI} and \textit{Insight}-HXMT when the \textasciitilde2.5\,Hz modulation are detected by \textit{NICER} are also marked with red points. } 
  \label{fig:hid_2}
\end{figure*}

\begin{figure}
	\includegraphics[width=\columnwidth]{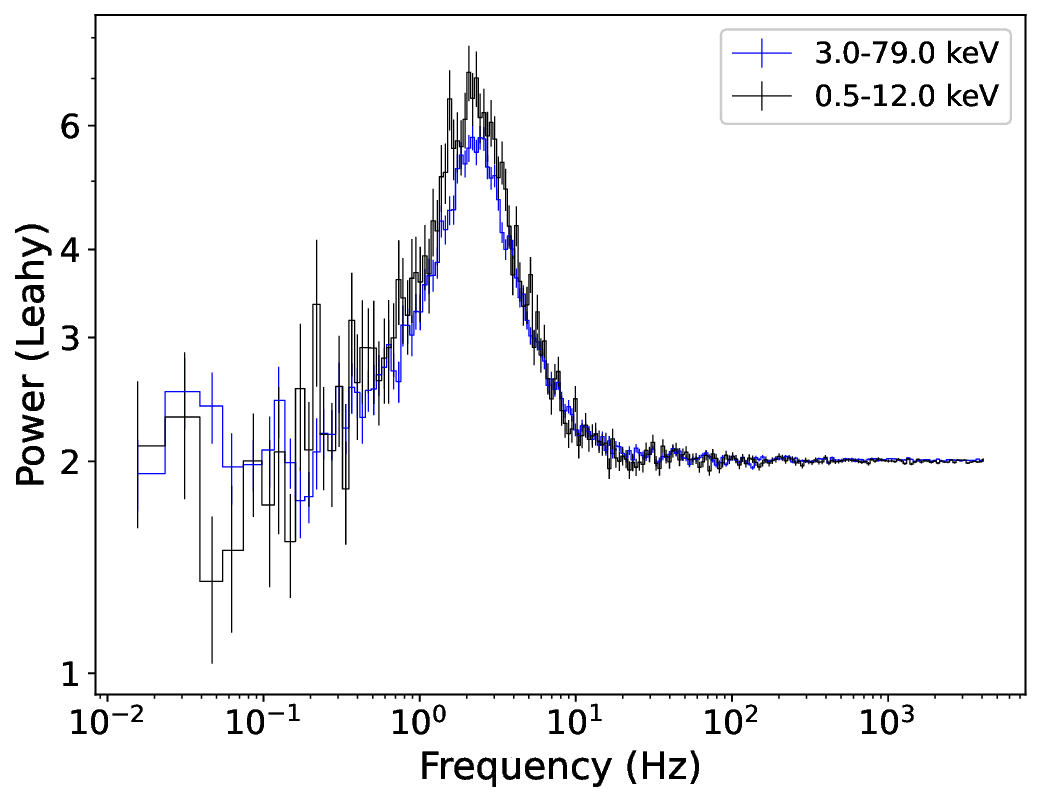}
	\includegraphics[width=\columnwidth]{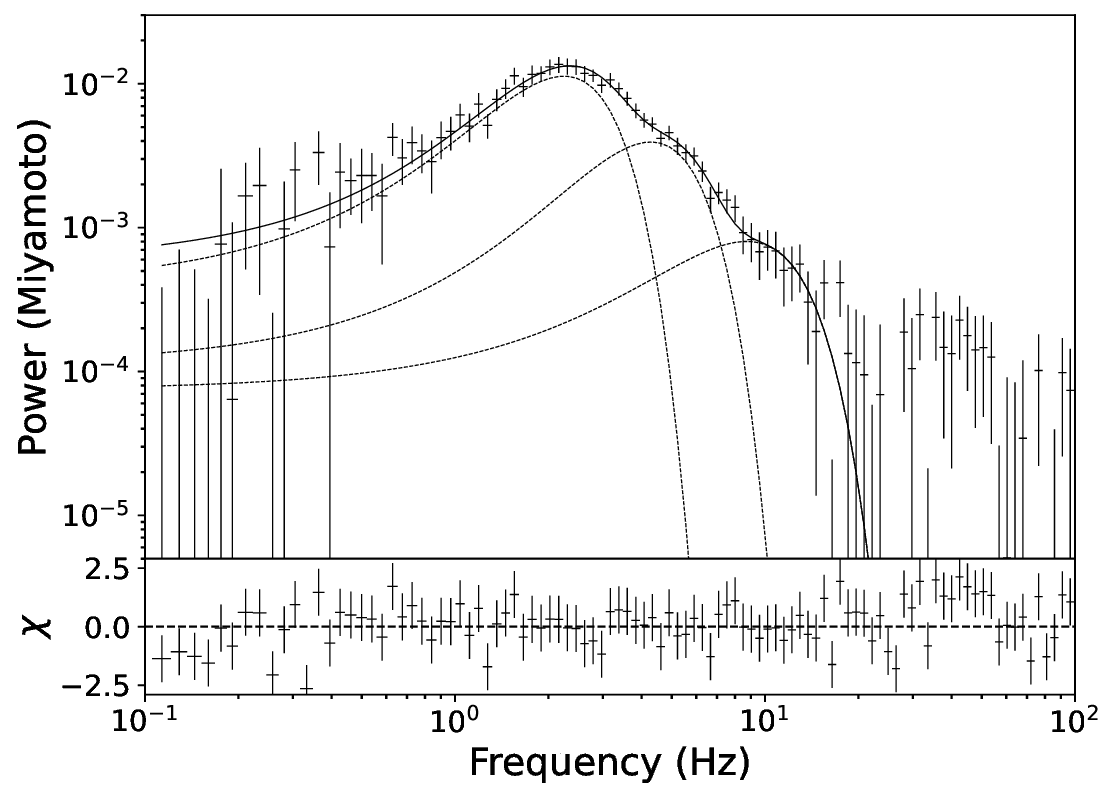}
    \caption{Upper panel: The Leahy-normalized power density spectra from \textit{NICER}’s ObsID 5533011601 (black; 0.5--12.0\,keV) and \textit{NuSTAR}’s dead time-corrected data (blue; 3--79\,keV) of MAXI J1816--195, respectively. The absence of a pulse peak is due to the rebinning of the frequency. Lower panel: The Miyamoto normalized power density spectrum of ObsID 5533011601 and the residuals of the PDS fitting. The black dashed line is the Gaussian model used for fitting. The large difference between the PSDs in the lower and upper panels is due to the Poisson noise background subtraction.}
  \label{fig:PDS_3}
\end{figure}

\begin{figure}
	\includegraphics[width=\columnwidth]{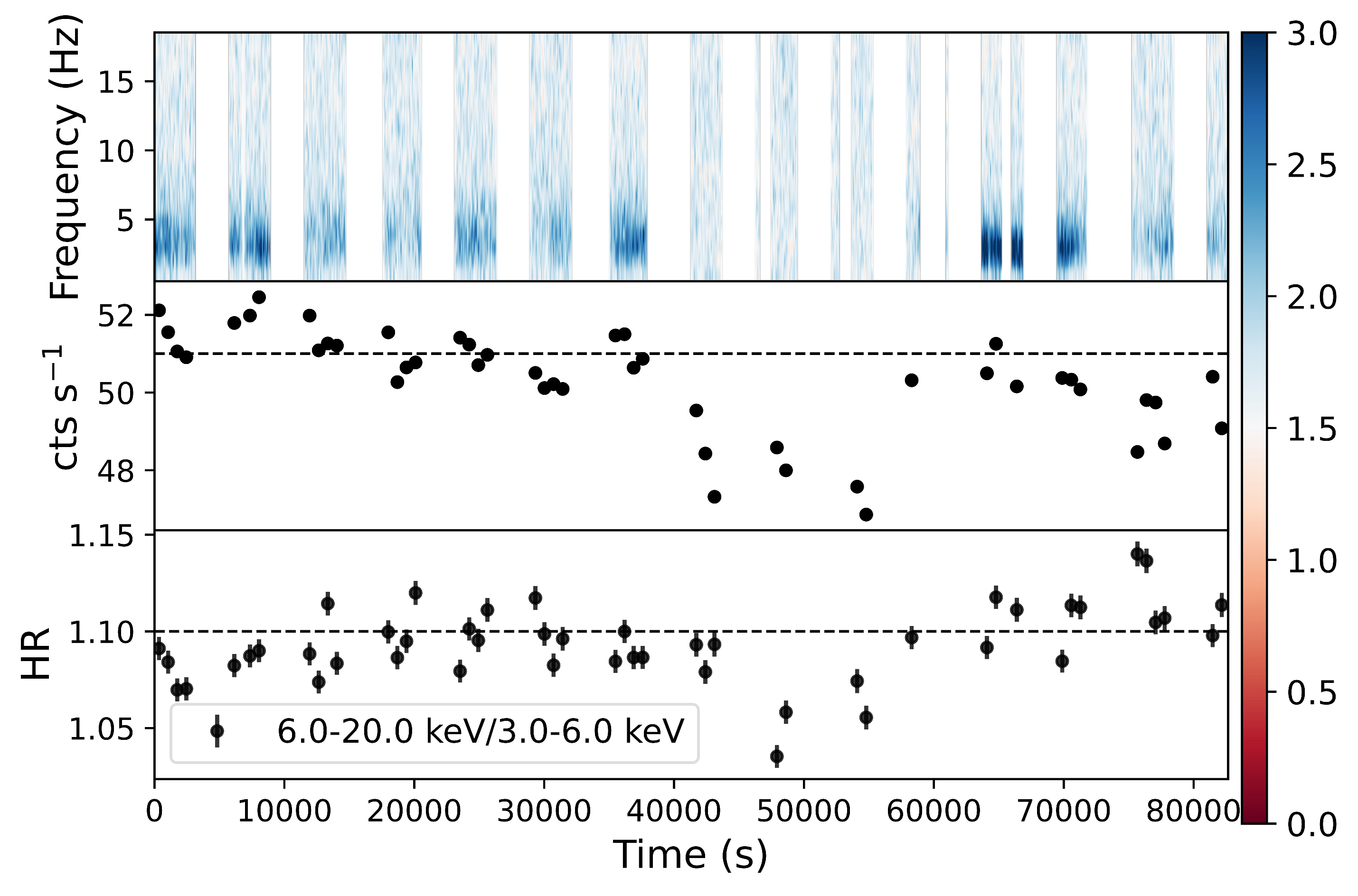}
	\includegraphics[width=\columnwidth]{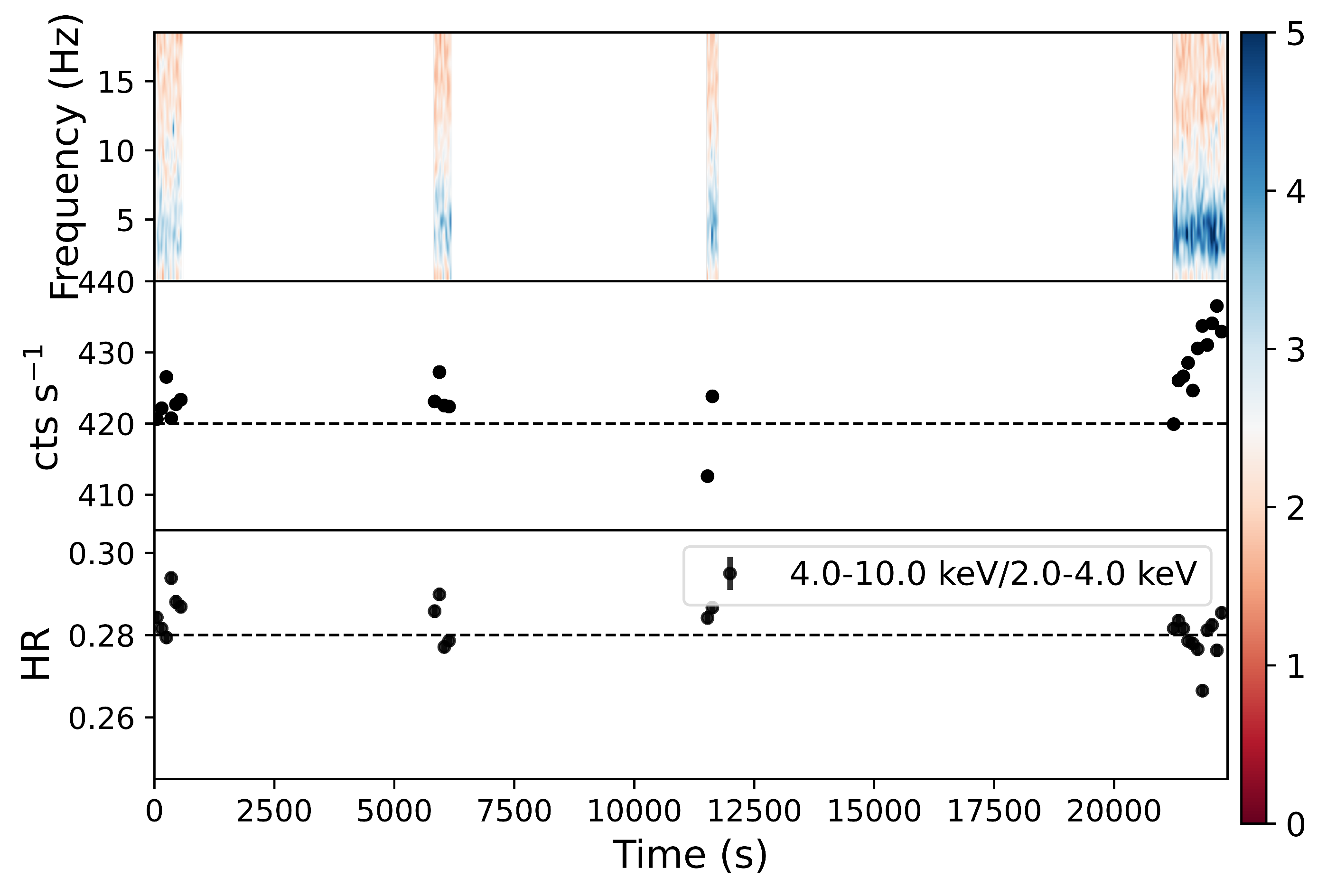}
	
    \caption{Upper panel: From top to bottom are the dynamic power spectrum, count rate (700\,s bins) and the hardness ratio (HR; 700\,s bins) of the 6–20\,keV to the 3–6\,keV obtained from \textit{NuSTAR} data. The dynamic power spectrum and count rate cover the 3--79 keV energy range. Lower panel: The same display for the ObsID 5533011101 is observed by \textit{NICER}. Dynamic power spectrum and count rate (100\,s bins) in the 0.5--12.0 keV energy band. HR (100\,s bins ) is 4--10\,keV versus 2--4\,keV. The right color bars represent the intensity of the Leahy-normalized dynamic power spectrum.}
  \label{fig:PDS_lc_5}
\end{figure}




\begin{table}
    \centering
    \caption{\textit{NICER} and \textit{NuSTAR} Observations of MAXI J1816--195 used in this paper. The observations with the \textasciitilde2.5\,Hz modulation are highlighted in bold text. Simultaneous observations by \textit{NICER} and \textit{NuSTAR} are marked in blue.}
	\label{tab:obsid_1}
    \begin{tabular}{@{}ccccc@{}}
    \hline
        Instrument & ObsIDs & Start time & Exposure time & Number of \\&&(MJD) & (s)& type-I bursts \\ \hline
        \textit{NICER} & 5202820102 & 59738.32 & 1210 & 0 \\ 
        \textit{NICER} & 5533010101 & 59738.77 & 2325 & 1 \\ 
        NICER & 5533010102 & 59739.35 & 3127 & 1 \\ 
        \textit{NICER} & 5533010103 & 59740.02 & 9177 & 2 \\ 
        \textit{NICER} & 5533010104 & 59741.61 & 2201 & 1 \\ 
        \textit{NICER} & 5533010105 & 59742.25 & 3869 & 1 \\ 
        \textit{NICER} & 5533010106 & 59743.09 & 10797 & 1 \\ 
        \textit{NICER} & 5533010107 & 59744.01 & 3441 & 0 \\ 
        \textit{NICER} & 5533010801 & 59746.14 & 4987 & 0 \\ 
        \textbf{\textit{NICER}} &  \textbf{5533010901} & \ \textbf{59747.16} &  \textbf{5099} &  \textbf{3} \\ 
        \textit{NICER} & 5533010108 & 59747.82 & 675 & 0 \\ 
        \ \textbf{\textit{NICER}} &  \textbf{5533010902} &  \textbf{59748.01} &  \textbf{1443} &  \textbf{0} \\ 
         \textbf{\textit{NICER}} &  \textbf{5533011001} &  \textbf{59748.15} &  \textbf{5504} &  \textbf{0} \\ 
         \textbf{\textit{NICER}} &  \textbf{5533011101} &  \textbf{59749.18} &  \textbf{2493} &  \textbf{0} \\ 
         \textbf{\textit{NICER}} &  \textbf{5533011201} &  \textbf{59750.13} &  \textbf{2330} &  \textbf{0} \\ 
         \textbf{\textit{NICER}} &  \textbf{5533011301} &  \textbf{59751.75} &  \textbf{1988} &  \textbf{1} \\ 
         \textbf{\textit{NICER}} &  \textbf{5533011401} &  \textbf{59752.65} &  \textbf{1712} &  \textbf{0} \\ 
         \textbf{\textit{NICER}} &  \textbf{5533011501} &  \textbf{59753.30} &  \textbf{5213} &  \textbf{0} \\
         \textbf{\color{blue}{ \textit{NuSTAR}}} &  \textbf{\color{blue}{90801315001}} &  \textbf{\color{blue}{59753.45}} &  \textbf{\color{blue}{35699}} &  \textbf{\color{blue}{4}} \\
         \textbf{\textit{\color{blue}{NICER}}} &  \textbf{\color{blue}{5533011601}} &  \textbf{\color{blue}{59754.15}} &  \textbf{\color{blue}{2373}} &  \textbf{\color{blue}{2}} \\ 
        \textit{NICER} & 5533011502 & 59754.92 & 613 & 0 \\ 
        \textit{NICER} & 5533011503 & 59755.24 & 1263 & 0 \\ 
        \textit{NICER} & 5533011701 & 59755.50 & 1653 & 0 \\ 
        \textit{NICER} & 5533011801 & 59756.02 & 1236 & 0 \\ 
        \textit{NICER} & 5533012101 & 59759.29 & 1131 & 0 \\ 
        \textit{NICER} & 5533012201 & 59760.72 & 2516 & 0 \\ 
        \textit{NICER} & 5533012301 & 59761.16 & 3146 & 0 \\ 
          \hline
    \end{tabular}
\end{table}

\subsection{\textit{Insight}-HXMT} 
\label{sec:2.3}
The hard X-ray Modulation Telescope (HXMT, also named \textit{Insight}-HXMT) is China’s first X-ray astronomy satellite with a broad energy in 1--250\,keV \citep[]{zhang2020}. Following the trigger of \textit{MAXI}, \textit{Insight}-HXMT began monitoring MAXI J1816--195 on June 8, 2022, and observations stopped on July 5, 2022. These observations covered the peak and the decay phase of the outburst. All observations are analyzed using the \textit{Insight}-HXMT processing software {\tt HXMTDAS v2.05}\footnote{\url{http://hxmtweb.ihep.ac.cn/software.jhtml}} according to standard processing procedures\footnote{\url{http://hxmtweb.ihep.ac.cn/SoftDoc/648.jhtml}}. Then, we generate light curves for the energy ranges of 2--10 keV (LE), 10--30 keV (ME), and 30--100 keV (HE), which do not include data between 30 s before the burst until 70 s after the burst. In addition, the arrival time of photons in the clean event file is corrected using the {\tt hxbary} tool. 

\section{Analysis and Results}
\label{sec:3}
\subsection{Light Curve, Hardness Ratio, and HIDs}
\label{sec:3.1}
The 2--10\,keV light curves for \textit{MAXI}/GSC\footnote{\url{http://maxi.riken.jp/top/index.html}}, \textit{NICER}, and \textit{Insight}-HXMT/LE observations are shown in the first three panels on the left of Fig.~\ref{fig:hr_lc_1}. The X-ray bursts have been removed from the \textit{NICER} and \textit{Insight}-HXMT/LE data. The light curves of the three satellites are consistent with regards to the evolution of the outburst from MJD 59738 to MJD 59761, which cover the peak and the decay phase of the outburst from MAXI J1816–195. The highest count rate of this source is \textasciitilde 480 counts s$^{-1}$, observed by \textit{NICER} on June 10, 2022, and then gradually decayed to \textasciitilde 15 counts s$^{-1}$ (MJD 59761). It should be noted that around MJD 59760, the count rates of HXMT are approximately 30\% of the peak fluxes, while the count rates in the MAXI and NICER data are less than 10\%. This is because HXMT is a collimated telescope with a large field of view, and there may be some contamination sources present within the field of view. The yellow shade of the left image of Fig.~\ref{fig:hr_lc_1} indicates the \textit{NuSTAR}'s observation interval for MAXI J1816–195. From this figure, it is clear that \textit{NuSTAR} and \textit{NICER} have a quasi-simultaneous observation on 2022 June 24 during the outburst decay phase. This 5\,s binned light curve of 3--79\,keV using \textit{NuSTAR}/FPMA is shown in the first panel on the right image of Fig.~\ref{fig:hr_lc_1}, which shows four bursts. The average count of persistent emissions is \textasciitilde 50 counts s$^{-1}$, and the count rate at bursts is as high as 26 times the average count. Its subgraph shows the burst profile with a 5\,s binned light curve and a hardness ratio. 

We define a hardness ratio (HR) from the ratio of the 4–10\,keV to the 2--4\,keV count rates in \textit{NICER}, and a hardness ratio for the 6--20\,keV to 3--6\,keV rates in \textit{NuSTAR}. The last panel on the left image of Fig.~\ref{fig:hr_lc_1} shows the HR evolution of the outburst using \textit{NICER} data. The HR first gradually decreases with the rise of the outburst count rates, maintains a relatively stable ratio of \textasciitilde 0.25 after reaching the peak, and then has a hardening bulge process at the red triangle mark (corresponding to the ObsIDs where the \textasciitilde2.5\,Hz modulation are detected; see detailed analysis in Section~\ref{sec:3.2}), and finally continues to harden as the outburst decays. In the subgraph on the right image of Fig.~\ref{fig:hr_lc_1}, during the decay phase of the burst, low-energy photons exhibit a long tail due to slow decay. The HR decreases and then returns to pre-burst levels.

Combining the intensity and HR, we see a counterclockwise evolution in Fig.~\ref{fig:hid_2}. For comparison, we also plot the hardness-intensity diagram (HID) of \textit{MAXI}/GSC and \textit{Insight}-HXMT/LE in the same energy band. As mentioned above, although the contamination sources may have some impact on the HXMT measurements in Fig.~\ref{fig:hid_2}, the evolutionary trend still holds. Noticeably, the high-statistics NICER data reveals a distinct hardened bulge at the red circle mark, which corresponds to the detection of the \textasciitilde2.5\,Hz modulation.

\begin{figure}
	\includegraphics[width=\columnwidth]{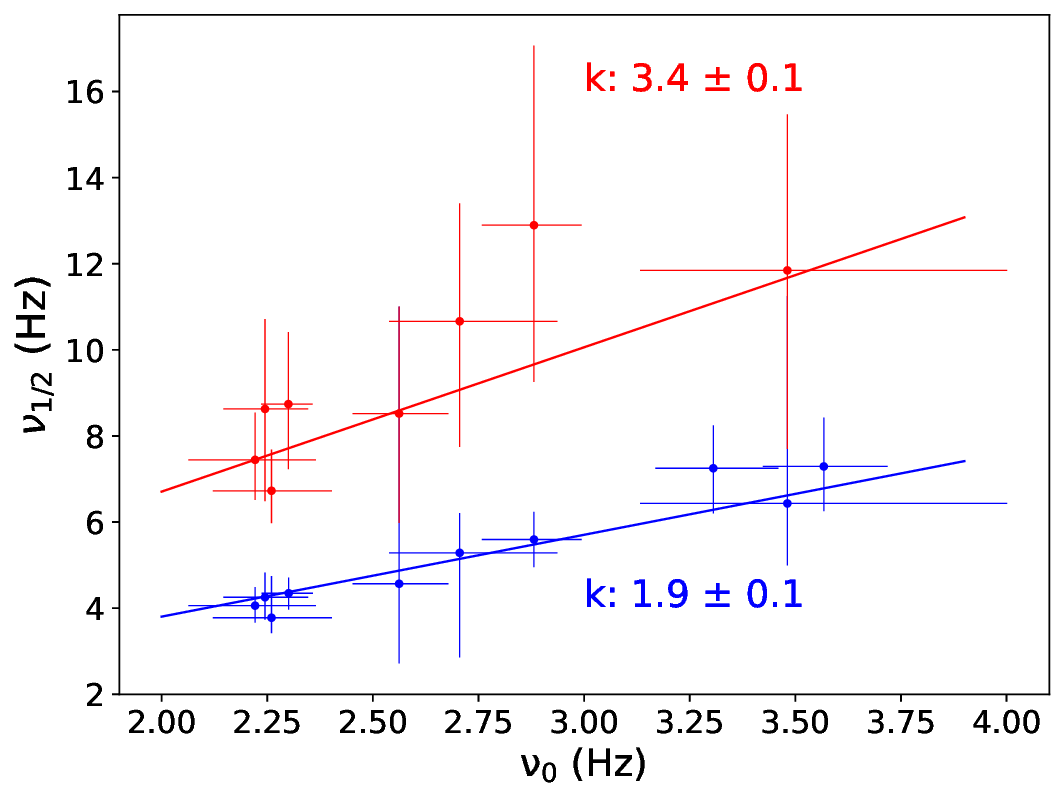}
	
    \caption{Correlations between the frequencies ($\nu_{1}$: blue; $\nu_{2}$: red) of the power spectral components plotted vs. $\nu_{0}$. The solid line is best fitted by $\nu_{1}$ (or $\nu_{2}$) = $k$$\nu_{0}$.}
  \label{fig:f2_f1_4}
\end{figure}

\begin{figure}
	\includegraphics[width=\columnwidth]{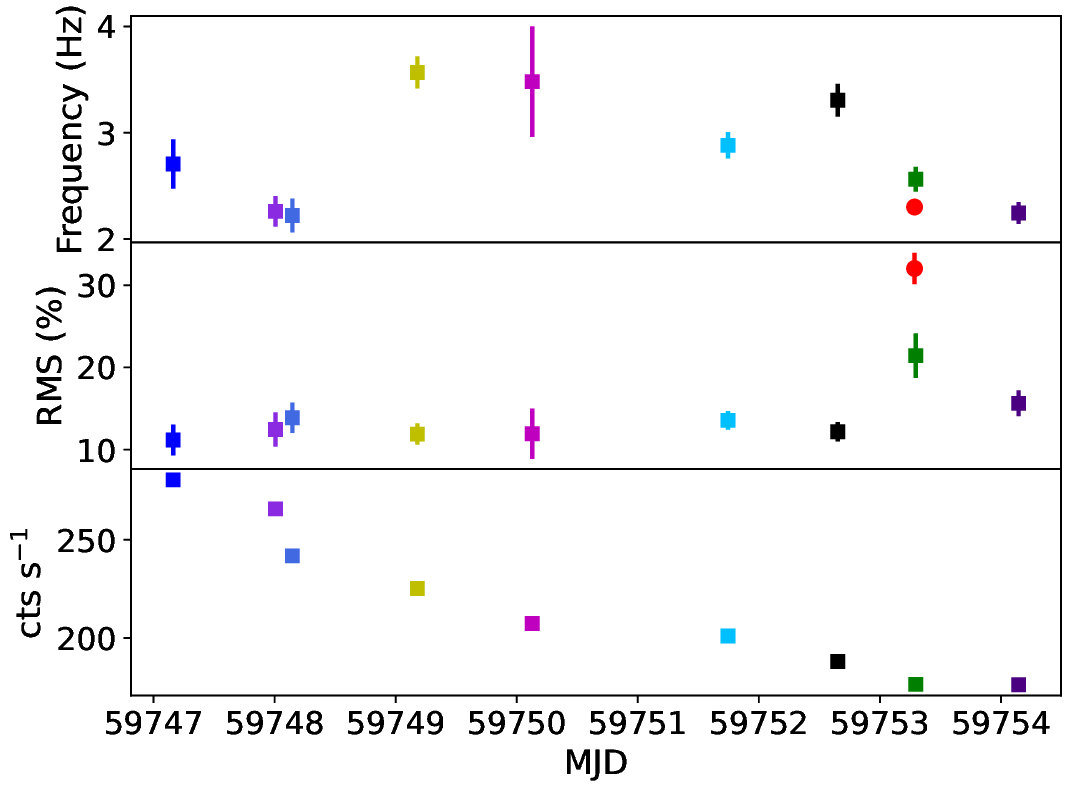}
	
    \caption{Centroid frequency (upper panel), fractional rms (middle panel) and counts rate (lower panel) versus time, which are from nine \textit{NICER} observations in the 0.5--12\,keV energy band (solid squares) and one \textit{NuSTAR} observation in the 3--79\,keV energy (solid circles). Different observations are represented by points with different colors.}
  \label{fig:f_rmd_lc_6}
\end{figure}

\begin{figure}
	\includegraphics[width=\columnwidth]{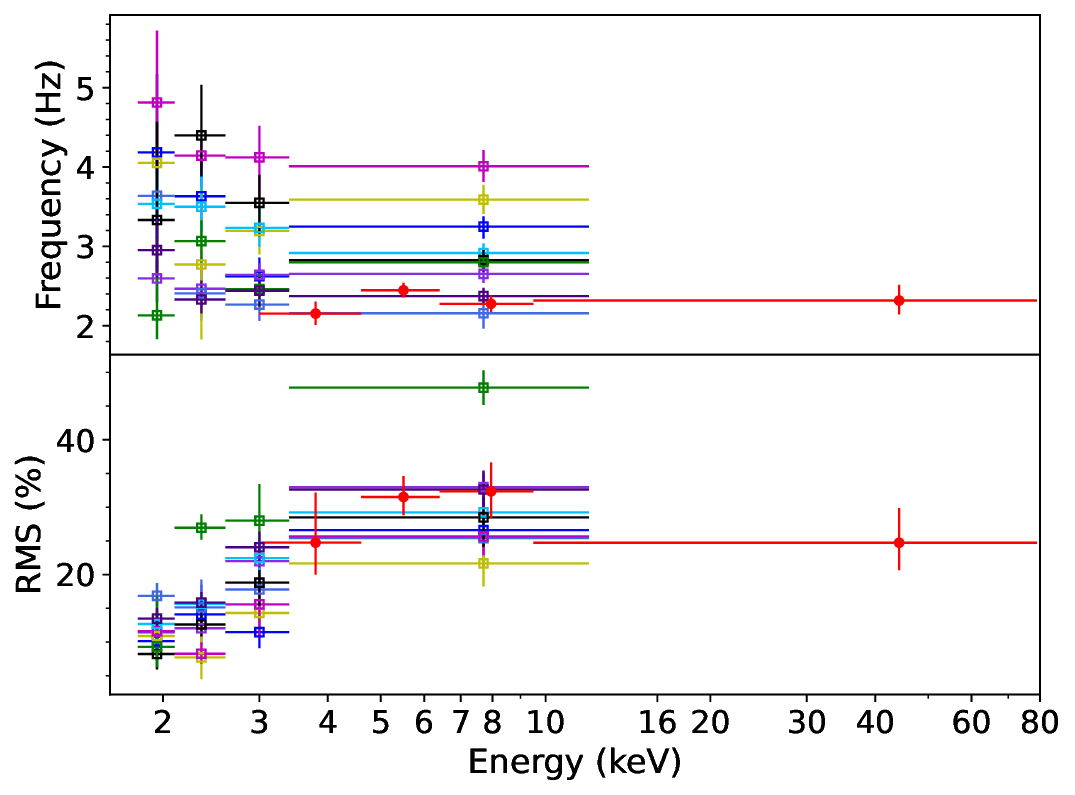}
	
    \caption{Upper panel: energy dependence of the centroid frequency of the \textasciitilde 2.5\,Hz modulation in MAXI J1816--195. Lower panel: rms spectra of the \textasciitilde2.5\,Hz modulation. Hollow squares of different colors in the same energy range represent nine different observations of \textit{NICER}, and solid red circles represent the \textit{NuSTAR} observation. The colors correspond to those used in Fig.~\ref{fig:f_rmd_lc_6}.}
  \label{fig:rms_energy_7}
\end{figure}

\begin{table*}
    \centering
    \caption{Multiple Gaussian fitting parameters for the PDS from \textit{NICER} and \textit{NuSTAR}. $\nu$, rms and Q represent the centroid frequency, fractional rms amplitude and quality factor, respectively. Subscript numbers indicate different Gaussian components.} 
    \renewcommand\arraystretch{1.37}
    \setlength{\tabcolsep}{1.5mm}{
	\label{tab:obsid_qpo_2}
    \begin{tabular}{@{}cccccccccccccc@{}}
    \hline
        ObsID & $\nu$$_{0}$ & rms$_{0}$ (\%) & Q$_{0}$  & $\nu$$_{1}$ & rms$_{1}$ (\%) & Q$_{1}$  & $\nu$$_{2}$ & rms$_{2}$ (\%) & Q$_{2}$  &$\chi^2$/d.o.f \\ \hline
        5533010901 & $2.7^{+0.2}_{-0.2}$ & $11.2^{+1.9}_{-1.7}$ & $2.3^{+0.4}_{-0.3}$  & $5.3^{+0.9}_{-2.4}$ & $10.4^{+2.2}_{-2.9}$ & $2.7^{+0.8}_{-1.5}$  & $10.7^{+2.7}_{-2.9}$ & $7.7^{+5.2}_{-1.7}$ & $2.4^{+0.6}_{-0.5}$  & 83/86\\ 
        5533010902 & $2.3^{+0.1}_{-0.1}$ & $12.5^{+2.1}_{-1.8}$ & $2.9^{+0.4}_{-0.4}$  & $3.8^{+1.0}_{-0.4}$ & $12.2^{+1.9}_{-2.1}$ & $3.0^{+1.0}_{-0.7}$  & $6.7^{+1.0}_{-0.8}$ & $12.9^{+1.3}_{-1.4}$ & $1.7^{+1.0}_{-1.0}$  &132/101\\ 
        5533011001 & $2.2^{+0.1}_{-0.2}$ & $13.9^{+1.8}_{-1.8}$ & $2.6^{+0.4}_{-0.4}$  & $4.1^{+0.4}_{-0.4}$ & $14.2^{+1.8}_{-1.8}$ & $2.8^{+0.5}_{-0.5}$  & $7.4^{+1.1}_{-0.9}$ & $12.3^{+1.8}_{-1.6}$ & $2.2^{+0.8}_{-0.8}$  &126/86\\ 
        5533011101 & $3.6^{+0.2}_{-0.1}$ & $11.9^{+1.3}_{-1.3}$ & $2.1^{+0.2}_{-0.3}$  & $7.3^{+1.1}_{-1.1}$ & $11.4^{+1.6}_{-1.5}$ & $1.9^{+0.4}_{-0.4}$  & ~.~.~. & ~.~.~. & ~.~.~.  &107/101\\ 
        5533011201 & $3.4^{+0.5}_{-0.3}$ & $11.9^{+2.7}_{-3.1}$ & $2.2^{+0.5}_{-0.5}$  & $6.4^{+4.8}_{-1.4}$ & $9.8^{+3.8}_{-3.6}$ & $2.8^{+6.1}_{-1.5}$  & $11.9^{+3.6}_{-4.1}$ & $7.1^{+3.7}_{-2.1}$ & $2.5^{+1.3}_{-1.1}$  &91/101\\ 
        5533011301 & $2.9^{+0.1}_{-0.1}$ & $13.6^{+1.1}_{-1.1}$ & $2.5^{+0.2}_{-0.2}$  & $5.6^{+0.6}_{-0.6}$ & $10.8^{+1.4}_{-1.4}$ & $2.6^{+0.5}_{-0.5}$  & $12.9^{+4.2}_{-3.6}$ & $7.4^{+2.1}_{-1.6}$ & $1.8^{+1.5}_{-1.5}$  &129/101\\ 
        5533011401 & $3.3^{+0.2}_{-0.1}$ & $12.2^{+1.1}_{-1.2}$ & $2.3^{+0.3}_{-0.3}$  & $7.3^{+2.5}_{-1.1}$ & $9.5^{+1.6}_{-1.4}$ & $2.5^{+0.7}_{-0.7}$  & ~.~.~. & ~.~.~. & ~.~.~.  &118/101\\ 
        5533011501 & $2.6^{+0.1}_{-0.1}$ & $21.4^{+2.4}_{-2.7}$ & $2.4^{+0.2}_{-0.3}$  & $4.6^{+6.4}_{-1.9}$ & $15.2^{+3.9}_{-3.6}$ & $2.6^{+3.6}_{-1.3}$  & $8.5^{+2.5}_{-2.5}$ & $13.4^{+4.1}_{-2.8}$ & $1.8^{+0.5}_{-0.5}$  &83/86\\ 
        5533011601 & $2.2^{+0.1}_{-0.1}$ & $15.6^{+1.5}_{-1.6}$ & $2.6^{+0.3}_{-0.3}$  & $4.3^{+0.6}_{-0.5}$ & $12.4^{+2.0}_{-2.2}$ & $2.7^{+0.6}_{-0.7}$  & $8.6^{+2.1}_{-2.1}$ & $8.7^{+2.6}_{-1.9}$ & $2.3^{+0.4}_{-0.3}$  &90/101\\ 
        90801315001 & $2.3^{+0.1}_{-0.1}$ & $32.1^{+1.8}_{-1.9}$ & $2.6^{+0.2}_{-0.2}$  & $4.4^{+0.4}_{-0.4}$ & $26.7^{+2.5}_{-2.5}$ & $2.4^{+0.4}_{-0.4}$  & $8.7^{+1.7}_{-1.5}$ & $22.1^{+3.41}_{-2.8}$ & $1.7^{+1.0}_{-1.0}$  &193/119\\ \hline
    \end{tabular}}
\end{table*}

\subsection{Timing Analysis}
\label{sec:3.2}
For timing analysis, the power density spectrum (PDS) calculated using {\tt POWSPEC} is written to text files. These files can be used as the input to {\tt flx2xsp} to create a "spectrum" file and diagonal response to read into {\tt XSPEC} for fitting (see \cite{2012Ingram} and references therein). In addition, \textit{NuSTAR} is primarily used for spectroscopy observation of dim sources, and there is a dead time of \textasciitilde 2.5\,ms for bright sources. This dead time can cause the resulting PDS to be severely distorted, so we used Fourier Amplitude Difference correction \citep[]{2018Bachetti} in {\tt Stingray} \citep[]{2022matteo} to correct the PDS. The error by $err=power/\sqrt N$, $N$ is the number of power averaged in each bin, which is added to each power and then imported {\tt XSPEC} for fitting.

We initially search for coherent signals by producing a PDS from the \textit{NICER} observation in the 0.5--12.0\,keV energy band. All individual power spectral estimates are averaged per observation (the type-I burst was removed) using 64\,s long segment and 1/8192\,s time bins, corresponding to a frequency resolution of 1/64\,Hz and a Nyquist frequency of 4096\,Hz. We perform Leahy normalization on the PDS \citep[]{1983Leahy} in order to assess the quality of the data depending on whether the poisson noise is 2. Ultimately, nine out of the twenty-six \textit{NICER}’s ObsIDs used in this paper show a strong and broad peak around 2.5\,Hz and do not observe any kHz QPOs. in the PDS (see  Table~\ref{tab:obsid_qpo_2} and Fig.~\ref{fig:PDS_3}). Such a signal is also found in \textit{NuSTAR}’s dead time corrected data in Fig.~\ref{fig:PDS_3}. Since the source is a bit dim, we did not find this peak in the PDS from \textit{Insight}-HXMT observations. The presence of a wide feature in the signal may suggest that the peaks evolve over time, so we generate dynamic power spectra for these observations with 0.03125 s bins, 16 s segments. As shown in Fig.~\ref{fig:PDS_lc_5}, the modulation is intermittent, and the central frequency of the modulation remains constant. We remove segments with no signal from each observation to reduce this intermittent broadening of the peak. We then regenerate PDSs, setting 1/512\,s time bins and applying Miyamoto normalization \citep[]{miyamoto1992}, to focus on peaks between 0.1--100\,Hz. We follow \citet{patruno2009} and \citet{patruno2013}'s analysis of SAX J1808.4--3658 and NGC 6440 X--2, and model these features in the PDS with two or three Gaussian functions in {\tt XSPEC} because the peak is quite broad and has a decline at lower frequencies. To assess the need for a third Gaussian component, we simulate $10^{5}$ spectra with the {\tt simftest} script, and if the significance of the third Gaussian exceeds 3$\sigma$, we include it in the fitting of the PDS. The results are shown in  Table~\ref{tab:obsid_qpo_2} and Fig.~\ref{fig:PDS_3}, a weak and insignificant broad component is observed around 40 Hz. This feature was detected in other observations, we thus do not give it much emphasis. 

The approximate harmonic relationships between the centroid frequencies of multiple Gaussians are $\nu_{1}$/$\nu_{0}$=$1.9\pm{0.1}$ and $\nu_{2}$/$\nu_{0}$=$3.4\pm{0.1}$, as shown in Fig.~\ref{fig:f2_f1_4}. The ratios based on $\nu_{1}$, however, gives $\nu_{0}$/$\nu_{1}$=$0.530\pm{0.01}$ and $\nu_{2}$/$\nu_{1}$=$1.88\pm{0.05}$, suggesting that the third component is also the fourth harmonics of the $\nu_{0}$ (or the second harmonics of $\nu_{1}$). This finding bears some resemblance to the centroid frequency ratios of simultaneous low-frequency QPOs in low-mass X-ray binaries 
\citep[e.g.][]{2020Doesburgh,2021Fei}. However, in their PDSs, both the fundamental and harmonic frequendcies are comparatively narrower, making them easier to be distinguished. In the case of MAXI J1816--195, these components are broader and weaker. Therefore, it is challenging to determine if all three components originate from genuine physical processes, requiring further observational research. Thus, this paper primarily focuses on discussing the strongest component ($\nu_{0}$), hereafter \textasciitilde2.5\,Hz modulation. Its quality factor ($Q$ = $\nu_{0}$/$FWHM$) is $> 2$. 

We calculate the fractional rms amplitude of this \textasciitilde2.5\,Hz modulation in order to study the strength of the low-frequency X-ray variability. The fractional rms amplitude of the Miyamoto normalized power density spectrum is defined as \citep[]{1989van,belloni1990}:

\begin{equation}
    rms=\sqrt{\int_{}^{} P(v) d v}.
	\label{eq:rms}
\end{equation}



We did not consider the background contribution because it is very low for both \textit{NICER} and \textit{NuSTAR}. As shown in Fig.~\ref{fig:f_rmd_lc_6}, the fractional rms of the \textit{NICER}’s nine observations are basically the same, and there is little correlation between the frequency, the fractional rms, and the source count rate. However, when considering short time scales ($<$1 Day), the presence of this modulation appears to be correlated with both the count rates and HRs, as demonstrated in Fig.~\ref{fig:PDS_lc_5}. Specifically, when the \textasciitilde2.5\,Hz modulation is observed, both the count rates and HRs exhibit higher values.

We also investigate the energy dependence of the \textasciitilde2.5\,Hz modulation rms, as well as the time lag between the energies. According to the equal count rate, each observation of \textit{NICER} is divided into six energy bands: 0.5--1.5\,keV, 1.5--1.8\,keV, 1.8--2.1\,keV, 2.1--2.6\,keV, 2.6--3.4\,keV, and 3.4--12.0\,keV. \textit{NuSTAR} is in four energy bands: 3.0--4.6\,keV, 4.6--6.4\,keV, 6.4--9.5\,keV, 9.5--79.0\,keV. The fractional rms amplitude increases with energy from \textasciitilde 12\% at 2\,keV to \textasciitilde 30\% at 8\,keV in Fig.~\ref{fig:rms_energy_7}. Above 5\,keV, the fractional rms remains approximately constant. It should be reminded that Fig.~\ref{fig:rms_energy_7} does not give the results of the low energy bands (0.5--1.5\,keV and 1.5--1.8\,keV) of \textit{NICER}. This is due to the very low significance of the modulation signals (Fig.~\ref{fig:PDS1.5}), which results in substantial uncertainties in the measurements.

\begin{figure}
	\includegraphics[width=\columnwidth]{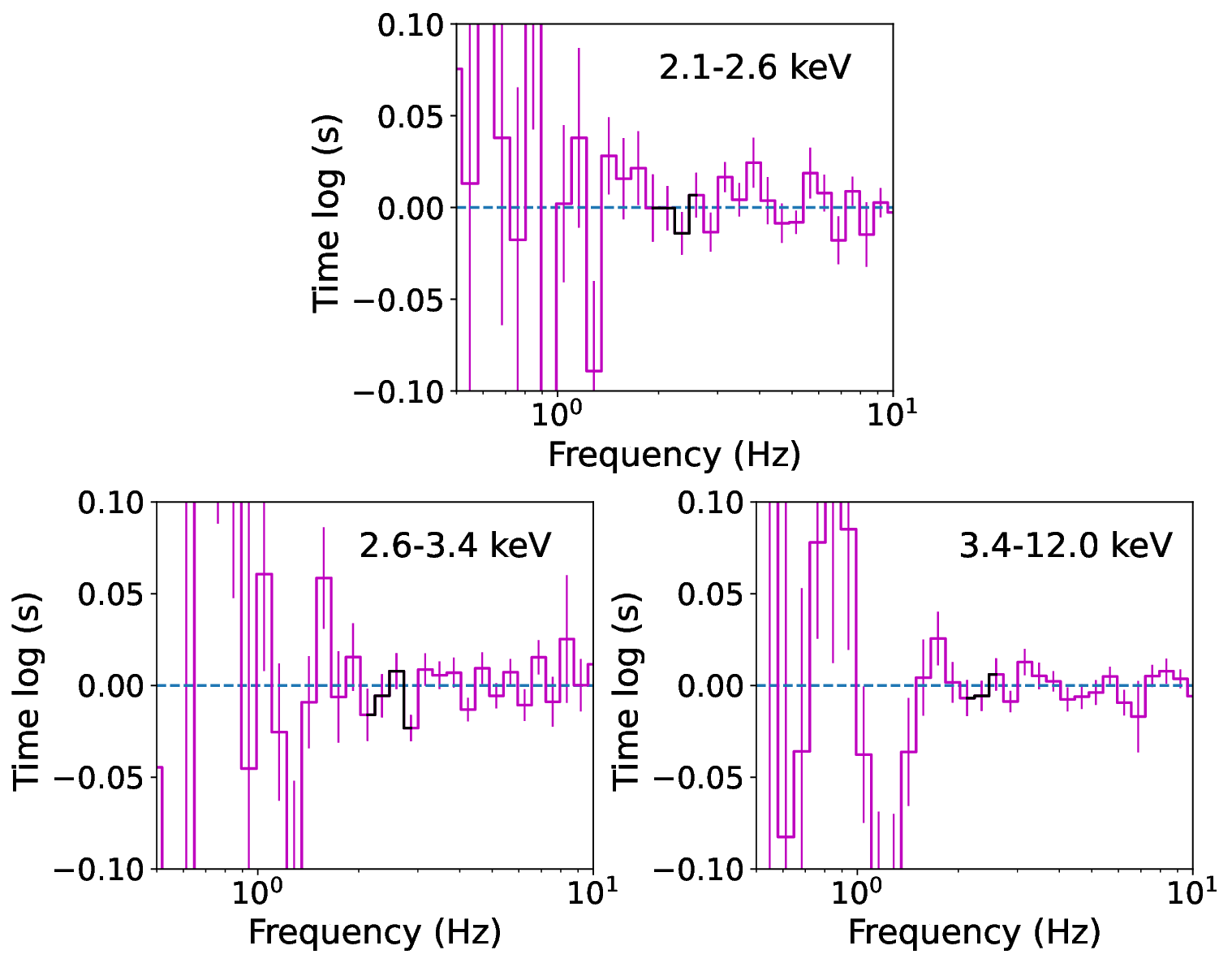}
	\includegraphics[width=\columnwidth]{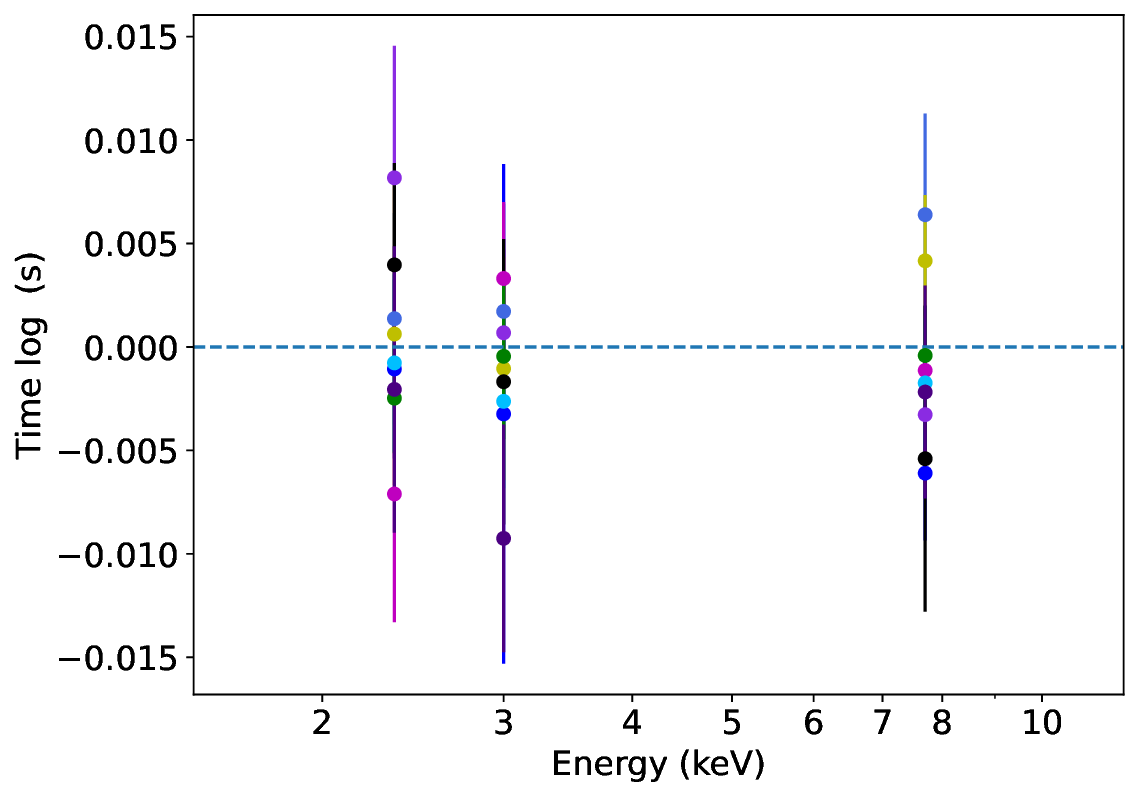}	
	
    \caption{Upper panel: An example (ObsID 5533011601) of the time lag-frequency spectra of the \textasciitilde2.5\,Hz modulation. The reference energy band is 1.8--2.1\,keV. The FWHM of the \textasciitilde2.5\,Hz modulation is highlighted by black. Lower panel: The time lag as a function of the energy of the \textasciitilde2.5\,Hz modulation from \textit{NICER}'s 9 ObsIDs. Colours are the same as in Fig.~\ref{fig:f_rmd_lc_6}. and Fig.~\ref{fig:rms_energy_7}.}
  \label{fig:time_lag_8}
\end{figure}

Following standard techniques in \citet{uttley2014x}, we initially calculate the lag-frequency spectrum using the AveragedCrossspectrum object in {\tt Stingray}. The upper panel in Fig.~\ref{fig:time_lag_8} corresponds to ObsID 5533011601 as an example, where the reference energy band is taken to be 1.8--2.1\,keV. The energy dependence of the time lags for all \textit{NICER}’s ObsIDs with the \textasciitilde2.5\,Hz modulation is obtained by averaging the lags within the FWHM of the \textasciitilde2.5\,Hz modulation in the lower panel of Fig.~\ref{fig:time_lag_8}. This result indicates that no time lags are detected.

\begin{figure}
	\includegraphics[width=\columnwidth]{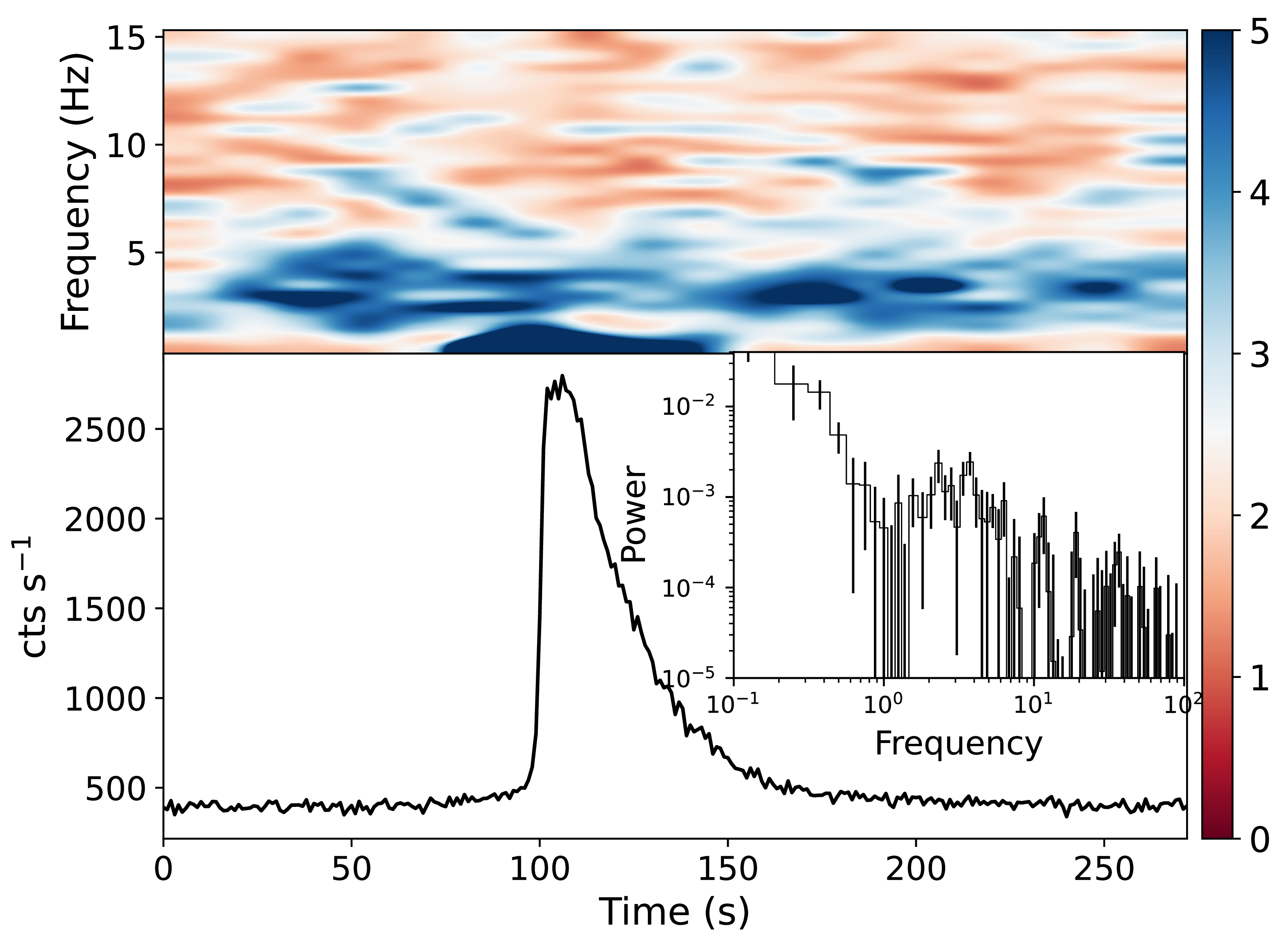}
	\includegraphics[width=\columnwidth]{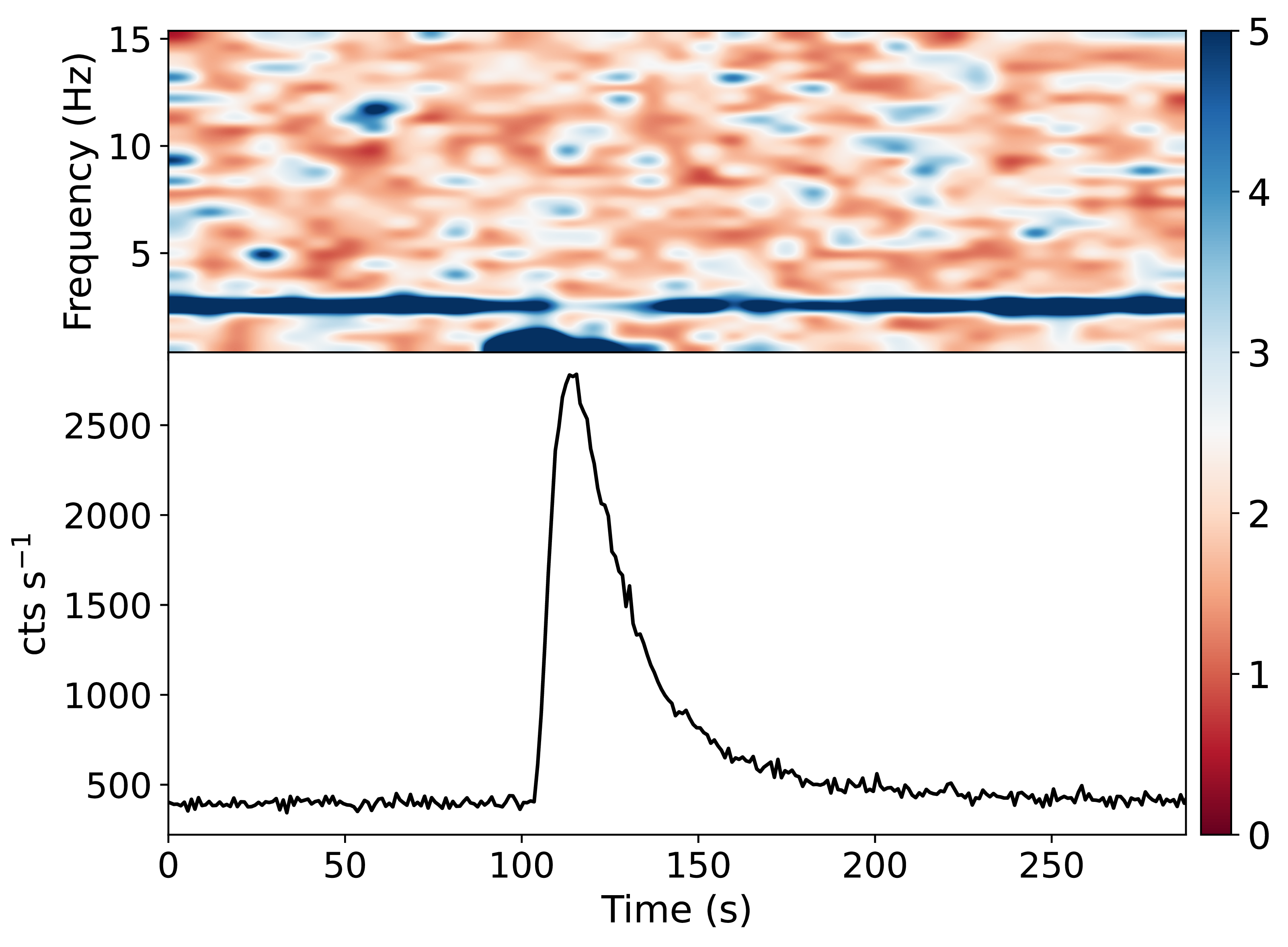}	
	
    \caption{Upper panel: The 1\,s light curve of the burst in ObsID 5533011301 observed by \textit{NICER}, and its corresponding dynamic power spectrum. The subplot is the power density spectrum from 100\,s to 150\,s on the timeline. Lower panel: The same plot for the simulated burst, which was created by the Inverse-Gamma distribution and the sinusoidal function, added with Poisson noise.}
  \label{fig:dps_burst_9}
\end{figure}

All of the above analyses are based on the MAXI J1816--195 persistent emission spectra and the type-I burst data are removed. In order to investigate whether there is a modulation signal during the burst, we generate dynamic power spectra by taking 300\,s of data starting 100\,s before the burst. We also extracted the first 50\,s data since the start of the burst to generate PDS. ObsID 5533011301 is shown as an example in the upper panel of Fig.~\ref{fig:dps_burst_9}, and it is not hard to see that the low-frequency modulation is not affected during the burst. The modulation in the subplot of the upper panel of Fig.~\ref{fig:dps_burst_9} is not smooth due to averaging together power spectra by using just three segments of data, but it does not affect the overall results. However, there is a serious noise at lower frequencies, which we conjecture is due to the Fourier transform being unfriendly to unstable signals, see \citet{polikar1996wavelet}. To prove our idea, we use an Inverse-Gamma distribution to simulate the type-I burst profile, followed by a sinusoidal function of 2.5\,Hz and Poisson noise. This simulated signal and its dynamic power spectrum are plotted in the lower panel of Fig.~\ref{fig:dps_burst_9}. The low-frequency strong noise corresponding to the position of the simulated burst on the figure proves our idea. In summary, the \textasciitilde2.5\,Hz modulation exists during the burst.

\begin{figure}
	\includegraphics[width=\columnwidth]{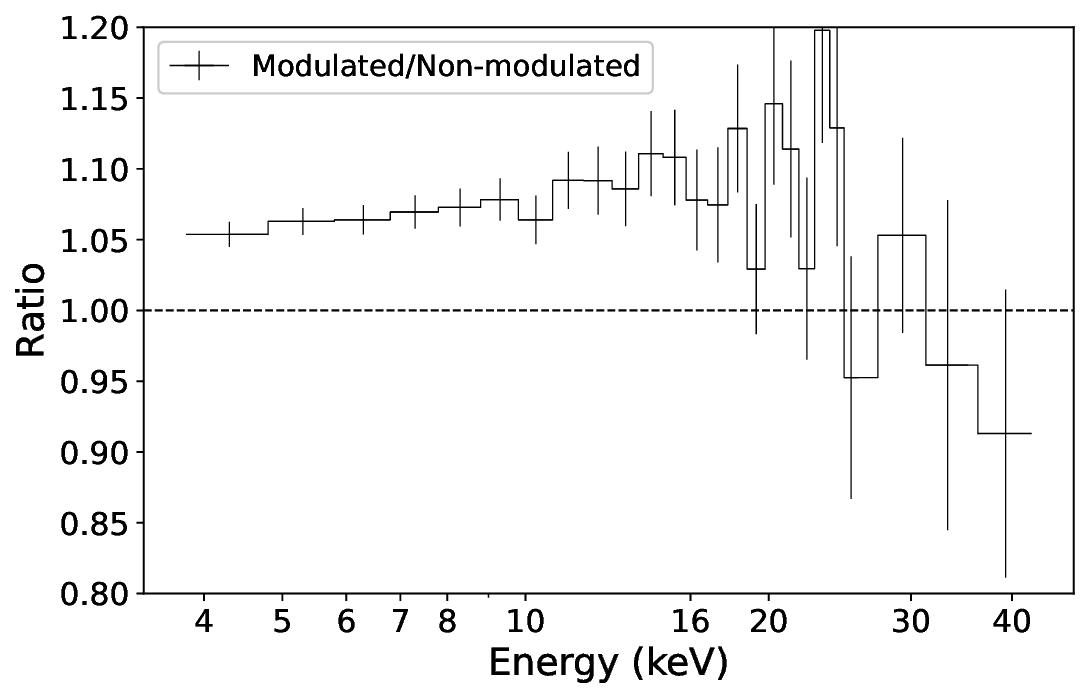}

    \caption{Data ratios between the periods with and without the \textasciitilde2.5\,Hz modulation from the \textit{NuSTAR}/FPMA observation.}
  \label{fig:ratio}
\end{figure}

\begin{figure}
	\includegraphics[width=\columnwidth]{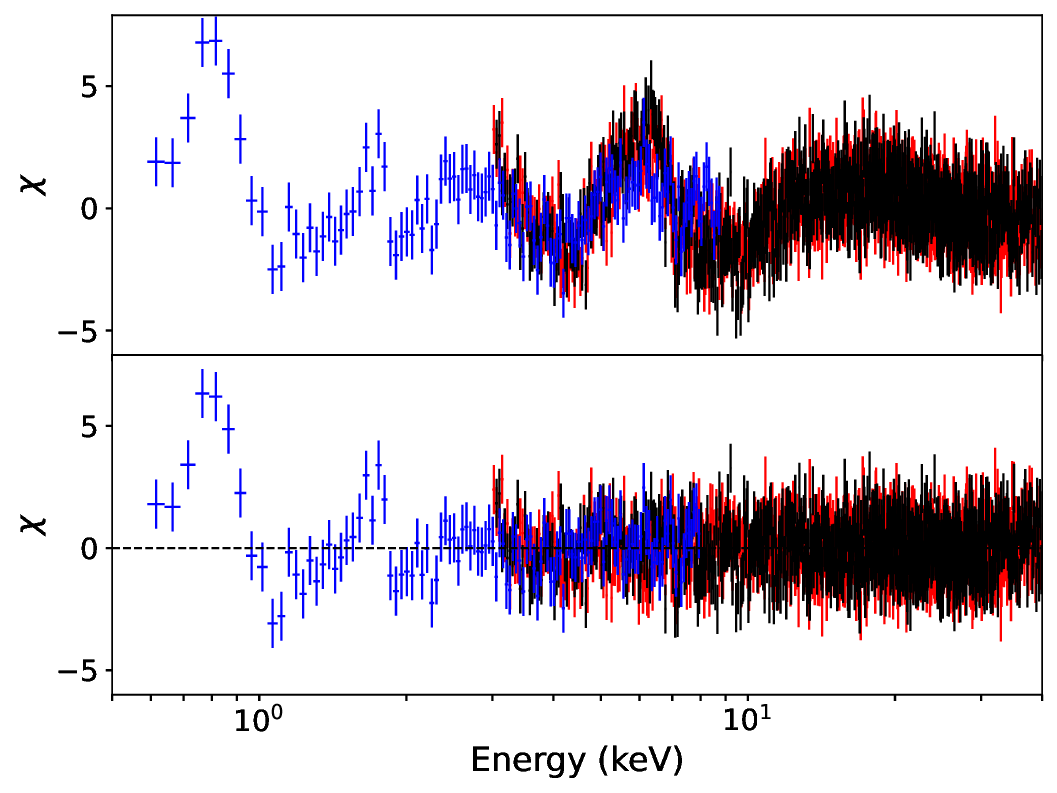}

    \caption{The residuals obtained with different models
in the range 0.5--40.0\,keV (blue for \textit{NICER}; red for FPMA; green for FPMB). Upper panel: {\tt const}$\times${\tt tbabs}$\times${(\tt nthcomp}+{\tt diskbb)}. Lower panel: {\tt const}$\times${\tt tbabs}$\times$ {\tt (relxillcp}$+${\tt diskbb)}}. 
  \label{fig:residuals_10}
\end{figure}

\subsection{Spectral Analysis}
\label{sec:3.3}
In this work, all spectra are fitted within {\tt XSPEC} v12.12.1. Abundances were set to WILM \citep[]{2000Wilms}, and cross-sections to VERN \citep[]{1996Verner}.

We conduct a persistent spectral analysis of MAXI J1816--195 using simultaneous observations from \textit{NICER} in the 0.5--10.0\,keV range and \textit{NuSTAR} (FPMA and FPMB) in the 3.0--40.0\,keV range on MJD 59754). We attempt to fit the data for periods with and without \textasciitilde2.5\,Hz modulation using the same model (model 1, as described below), but no significant variations in the spectral parameters were found. It is worth noting that for the spectra with modulation, the values of parameter $\Gamma$ = $1.90^{+0.01}_{-0.03}$ and $kT_{\rm e}$ = $9.39^{+1.99}_{-0.87}$\,keV, while for the spectra without modulation, the values of $\Gamma$ = $1.94\pm{0.01}$ and $kT_{\rm e}$ = $11.94^{+2.52}_{-1.23}$\,keV. However, considering the uncertainties, we cannot conclude that the spectra during the modulation period are harder with high significance. Therefore, we also plot the data ratios of periods with and without the \textasciitilde2.5\,Hz modulation from the \textit{NuSTAR}/FPMA observation in Fig~\ref{fig:ratio}. The ratios greater than one and slightly steeper with increasing energy are consistent with the timing analysis mentioned in Section~\ref{sec:3.2}, indicating that the fluxes become larger and the spectra are slightly harder when the \textasciitilde2.5\,Hz modulation is detected. But since the ratios do not exhibit a very strong dependence on energy, to enhance the signal-to-noise ratio, we subsequently studied the persistent emission of the entire \textit{NuSTAR} observation.


Following the procedures of \citet{2022Chen}, \citet{2022Mandal} and \citet{2022Bult}, we initially try to fit the joint \textit{NICER} and \textit{NuSTAR} spectra by an absorbed disk black-body ({\tt diskbb}) plus a thermally Comptonised continuum ({\tt nthcomp}) with seed photons from the accretion disc. A constant multiplication factor is included to account for calibration differences between different telescopes. Residuals with the broad emission feature in 6--7\,keV and excess in 10--30\,keV suggest a possible emission line from Fe-K and a Compton hump from the reflection of hard X-rays by the cool accretion disc. These features are evident in Fig.~\ref{fig:residuals_10} (upper panel). We therefore proceed by modeling our data with the self-consistent reflection model {\tt relxill v2.2} \footnote{\url{http://www.sternwarte.uni-erlangen.de/~dauser/research/relxill/index.html}}. The {\tt relxillCP} model, which describes the reflection using {\tt nthcomp} as an illuminating continuum, is used. We have assumed an unbroken emissivity profile ($\propto$ $r^{\rm -q}$) with a fixed slope of $q = 3$ \citep[]{2012Wilkins}. We also have fixed the outer radius $R_{\rm out}$ =1000\,$R_{\rm g}$, as the sensitivity of the reflection fit decreases with increasing outer disc radius, and set a redshift of 0. From measurements of the NS spin frequency with 528\,Hz \citep[]{2022Bult}, the dimensionless spin parameter $a = 0.248$  has been approximated using the relation $a = 0.47/P_{\rm ms}$ \citep[]{2000Braje}, where $P_{\rm ms}$ is the spin period in ms. We notice that the large residuals using the {\tt const}$\times${\tt tbabs}$\times${\tt (relxillcp}$+${\tt diskbb)} model with a $\chi^2$/dof = 1884/1683 = 1.12 are still at soft energies, as shown in the lower panel of Fig.~\ref{fig:residuals_10}; there are two edge-like shapes near \textasciitilde 0.9 and \textasciitilde 1.8 keV. These remaining features maybe come from \textit{NICER}’s calibration systematics\footnote{\url{https://heasarc.gsfc.nasa.gov/docs/nicer/analysis_threads/arf-rmf/}}\footnote{\url{https://heasarc.gsfc.nasa.gov/docs/nicer/analysis_threads/plot-ratio/}} (see e.g. \cite{2020Ludlam,2021Ludlam}). Therefore, two {\tt edge} models are added in the fitting, and the overall model becomes {\tt const}$\times${\tt edge}$\times${\tt edge}$\times${\tt tbabs}$\times${\tt (relxillcp}$+${\tt diskbb)}, hereafter Model 1. The addition of the two {\tt edge} model improves the overall fit significantly to $\chi^2$/dof = 1737/1678 = 1.03.

\begin{table}
	\centering
	\caption{Best-fitting spectral parameters of the \textit{NICER} and \textit{NuSTAR} observations of MAXI J1816--195 using Model 1: {\tt const}$\times${\tt edge}$\times${\tt edge}$\times${\tt tbabs}$\times${\tt (relxillcp}$+${\tt diskbb)}. Uncertainties are given at 90\%.}
	\renewcommand\arraystretch{1.36}
	\begin{tabular}{ccc}
			\hline
			Model & Parament (unit) & Value \\
			\hline
			TBABS & $N_{\rm H}\ (\times10^{22}\ {\rm cm^{-2}})$ & $2.27^{+0.04}_{-0.04}$ \\
			EDGE & $E_{\rm edge,1}  $ (keV)& $1.84^{+0.04}_{-0.04}$\\
			EDGE & $E_{\rm edge,2}$ (keV) & $0.90^{+0.02}_{-0.03}$\\
			DISKBB & $kT_{\rm in}\ ({\rm keV})$  & $0.55^{+0.01}_{-0.01}$  \\
			   & $N_{\rm diskbb}$  &$916^{+74}_{-62}$ \\
			  
			RELXILLCP & $\Gamma$ & $1.96^{+0.02}_{-0.01}$ \\
			 & $kT_{\rm e}\ ({\rm keV})$ & $10.06^{+0.40}_{-0.35}$ \\
			 & $R_{\rm in}$\ $(\times R_{\rm ISCO})$ & $1.14^{+0.09}_{-0.10}$\\
			 & $a$ & 0.248\ (fixed)\\
			 & Refl\_frac & $0.18^{+0.04}_{-0.02}$  \\
			 & $i\ (^{\circ})$ & $<13.23$  \\
			 & $\log\xi$ & $3.28^{+0.16}_{-0.25}$  \\
			 & $\log N\ ({\rm cm^{-3}})$ & $17.95^{+0.67}_{-0.83}$\\
			 & $A_{\rm Fe}$\ $(\times \rm solar)$ & $2.13^{+0.73}_{-0.84}$ \\
			 & Norm\ ($\times10^{-4}$) & $55^{+2}_{-3}$\\
			\hline
			CFLUX & $F_{\rm total}\ (\times10^{-9}\ {\rm ergs\ \rm s^{-1}\ \rm cm^{-2}})$ &$5.42^{+0.07}_{-0.07}$\\
			 (0.5--79\,keV)&  $F_{\rm diskbb}\ (\times10^{-9}\ {\rm ergs\ \rm s^{-1}\ \rm cm^{-2}})$ & $1.36^{+0.06}_{-0.06}$\\
			 & $F_{\rm relxillcp}\ (\times10^{-9}\ {\rm ergs\ \rm s^{-1}\ \rm cm^{-2}})$ &$4.05^{+0.02}_{-0.02}$\\
			\hline
			 & $\chi^2$/d.o.f & 1.03\ (1738/1678) \\
			\hline
	\end{tabular}
	\label{tab:3}
\end{table}

Regrettably, the downside of RELXILLCP is the fixed seed photon temperature of 0.01\,keV. In an attempt to mitigate this limitation, we explored the use of the {\tt nthratio} model \footnote{\url{https://github.com/garciafederico/nthratio}}. However, the revised parameters obtained from this model did not significantly impact the subsequent discussions. Moreover, considering the self-consistency of the reflection model, we decided not to include this modification in the paper. We look forward to further improvements and optimizations of future models.

\begin{figure}
	\includegraphics[width=\columnwidth]{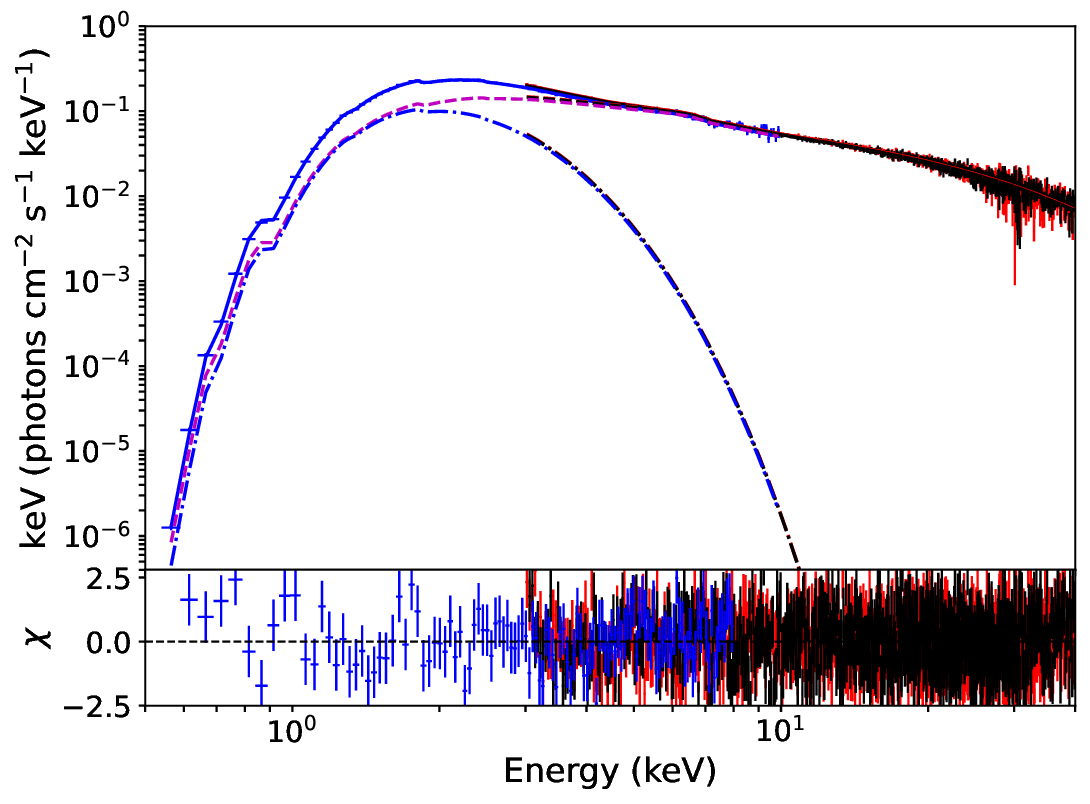}

    \caption{Spectra, model components, and spectral residuals for \textit{NICER} and \textit{NuSTAR} observations of MAXI J1816--195 with Model 1. The dash-dotted line shows the {\tt diskbb} model, while the dashed line represents the {\tt relxillcp} model.}
  \label{fig:spectrum_11}
\end{figure}

\begin{figure}
	\includegraphics[width=\columnwidth]{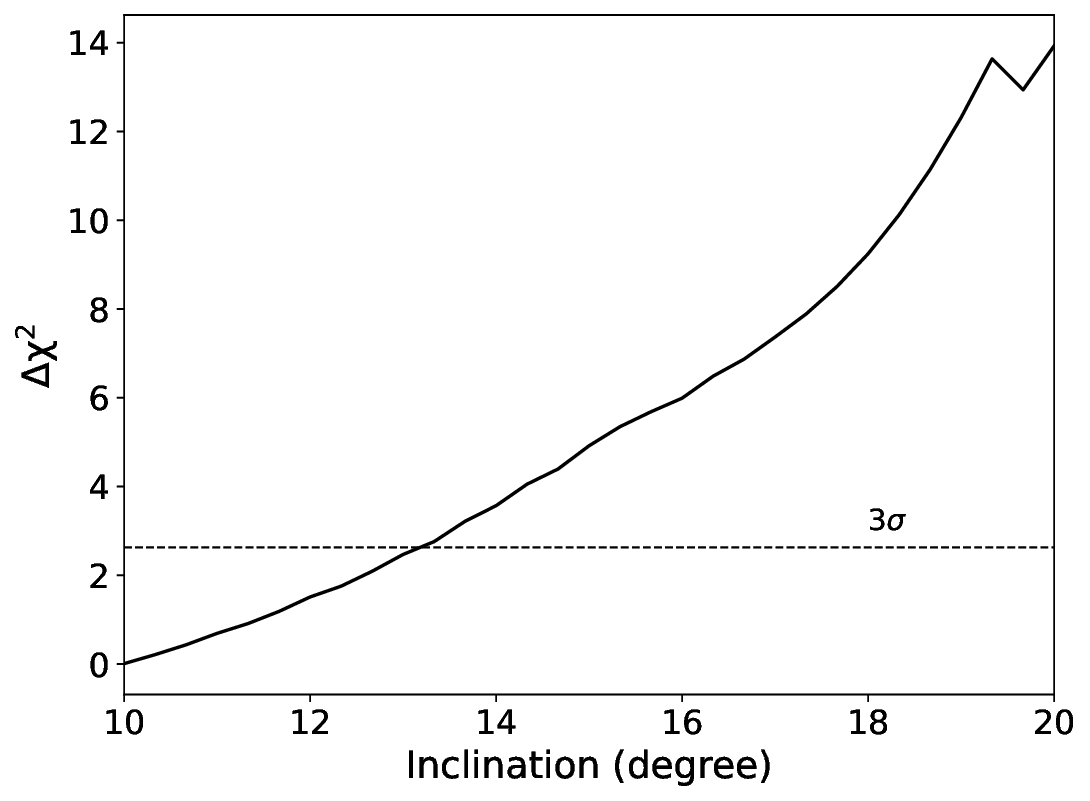}

    \caption{Variation of $\Delta \chi^{2}$ as a function of inclination obtained from the relativistic reflection model (Model 1). The horizontal line represents $3\sigma$ significance level.}
  \label{fig:inclination_12}
\end{figure}

The best-fitting parameters are listed in Table~\ref{tab:3} and the persistent spectra are shown in Fig.~\ref{fig:spectrum_11}. We obtain a hydrogen column density ($N_{\rm H}$) of $(2.27\pm{0.04}) \times 10^{22}$\ ${\rm cm^{-2}}$, an inner disc temperature of $kT_{\rm in}$ = $0.55\pm{0.01}$\,keV, a power-law photon index of $\Gamma$ = $1.96^{+0.02}_{-0.01}$, and the electron temperature $kT_{\rm e}$ = $10.06^{+0.40}_{-0.35}$\,keV. The iron abundance $A_{\rm Fe}$ in solar abundance and the ionization parameter $\log\xi$ of the disc are $2.13^{+0.73}_{-0.84}$ and $3.28^{+0.16}_{-0.25}$, respectively, which is consistent with other NS LMXBs (see e.g. \citet{2019Sharma}). It predicts a small inner disc radius of $R_{\rm in}$ = (1.04--1.23)\,$R_{\rm ISCO}$ from the reflection spectra. For a spinning NS with $a = 0.248$, $R_{\rm ISCO}$ can be approximated using $R_{\rm ISCO}$ = 6$R_{\rm g}$(1--0.54$a$) \citep[]{1998Miller}. An inner disc radius of $R_{\rm in}$ = (5.41--6.40)\,$R_{\rm g}$ = (11.20--13.24)\,km is measured for Model 1 ($R_{\rm g} = GM/c^{2}$ is the gravitational radii, which is 2.07\,km for M = 1.4\,$M_{\rm \odot}$). This value is roughly consistent with the inner disc radius ($<19$\,km) inferred from the {\tt diskbb} normalization $N_{\rm diskbb}$ = $(R_{\rm in,diskbb}/D_{\rm 10})^{2} \cos{i}$, where the upper limit of the distance is assumed to be 6.3\,kpc \citep[]{2022Chen} and the lower limit of inclination is $0^{\circ}$. If we consider the emission from the disk to be Comptonized by the corona, the inner radius of the disk would be slightly larger \citep{1998Kubota}. In addition, our spectral fits can only give the upper limit on inclination $< 13.23^{\circ}$ at $3\sigma$ confidence level. We compute $\chi^{2}$ for inclination using steppar in {\tt XSPEC}. The variation of $\chi^{2}$ of the fit versus the inclination ($i$) is shown in Fig.~\ref{fig:inclination_12}.


 In addition, by considering that the presence of a hard surface of a neutron star might imply a potential blackbody component, we also test the fits by adding a {\tt bbodyrad} to Model 1, which we define as Model 2 ({\tt const}$\times${\tt edge}$\times${\tt edge}$\times${\tt tbabs}$\times${\tt (relxillcp}$+${\tt bbodyrad}$+${\tt diskbb)}).
 We find that the blackbody component is not significant with a temperature $kT_{\rm bb}$ = 2.05\,keV and a radius $R_{\rm bb}$ = 0.52\,km, and the values of other parameters do not change (Table~\ref{tab:4}). Moreover, $\chi^{2}$ is just improved from 1737 to 1717. The weak blackbody component is similar to the 2017 outburst of IGR J16597--3704 \citep[]{2018Sanna} and the 2019 outburst of SAX J1808.4--3658 \citep[]{2022Sharma}. Therefore, the following discussion is based on the parameters obtained from Model 1.





\begin{table}
	\centering
	\caption{Best-fitting spectral parameters of the \textit{NICER} and \textit{NuSTAR} observations using Model 2: {\tt const}$\times${\tt edge}$\times${\tt edge}$\times${\tt tbabs}$\times${\tt (relxillcp}$+${\tt bbodyrad}$+${\tt diskbb)}. Uncertainties are given at 90\%.}
	\renewcommand\arraystretch{1.36}
	\begin{tabular}{ccc}
			\hline
			Model & Parament (unit) & Value \\
			\hline
			TBABS & $N_{\rm H}\ (\times10^{22}\ {\rm cm^{-2}})$ & $2.20^{+0.04}_{-0.06}$ \\
			EDGE & $E_{\rm edge,1} $(keV)& $1.85^{+0.03}_{-0.03}$\\
			EDGE & $E_{\rm edge,2} $ (keV) & $0.88^{+0.03}_{-0.03}$\\
			DISKBB & $kT_{\rm in}\ ({\rm keV})$  & $0.59^{+0.02}_{-0.02}$  \\
			   & $N_{\rm diskbb}$  &$736^{+66}_{-70}$ \\
			BBODYRAD & $kT_{\rm bb}\ ({\rm keV})$& $2.05^{+0.57}_{-0.25}$\\
			 &$N_{\rm bb}$&$0.69^{+1.12}_{-0.22}$\\
			RELXILLCP & $\Gamma$ & $1.88^{+0.04}_{-0.08}$ \\
			 & $kT_{\rm e}\ ({\rm keV})$ & $10.27^{+0.57}_{-0.56}$ \\
			 & $R_{\rm in}$\ $(\times R_{\rm ISCO})$ & $< 1.12$\\
			 & $a$ & 0.248\ (fixed)\\
			 & Refl\_frac & $0.38^{+0.13}_{-0.20}$  \\
			 & $i\ (^{\circ})$ & $< 13.81$  \\
			 & $\log\xi$ & $3.64^{+0.36}_{-0.27}$ \\
			 & $\log N\ ({\rm cm^{-3}})$ & $< 18.12$\\
			 & $A_{\rm Fe}$\ $(\times \rm solar)$ & $2.27^{+4.41}_{-1.14}$ \\
			 & Norm\ ($\times10^{-4}$) & $38^{+3}_{-13}$\\
			\hline
			 & $\chi^2$/d.o.f & 1.02\ (1717/1676) \\
			\hline
	\end{tabular}
	\label{tab:4}
\end{table}

\section{Discussion}
\label{sec:4}
In this paper, we have analyzed the spectral and timing properties of the newly discovered millisecond pulsar MAXI J1816--195 using \textit{NICER} and \textit{NuSTAR}, and find a strong low-frequency modulation in the 2--4\,Hz range during the decay stage of the outburst. The modulation is well described by multiple Gaussians. The strongest component is the main subject, called \textasciitilde2.5\,Hz modulation. The broadband X-ray spectra of MAXI J1816--195 were well-fit by a combination of the reflection model and a disk black-body component over the range of 0.5 to 40\,keV.

\subsection{The properties of MAXI J1816--195 from the reflection model}
The persistent spectra of MAXI J1816--195 during simultaneous observations of \textit{NICER} and \textit{NuSTAR} are well-fitted by Model 1 ({\tt const}$\times${\tt edge}$\times${\tt edge}$\times${\tt tbabs}$\times${\tt (relxillcp}$+${\tt diskbb)}). According to Model 1, the flux in 0.5--79\,keV is $(5.42\pm{0.07})\times10^{-9}\ {\rm ergs\ \rm s^{-1}\ \rm cm^{-2}}$, corresponding to 14.3\%$L_{\rm Edd}$ at a distance of 6.3\,kpc for $L_{\rm Edd}$ = $1.8\times10^{38}\ {\rm ergs\ \rm s^{-1}}$ assuming a neutron star mass of 1.4 $M_{\odot}$. Equation (2) in \citet{2008Galloway} is used to estimate the mass accretion rate of MAXI J1816--195. We assume $c_{\rm bol} = 1.38$, $1+z = 1.31$ for a NS with mass ($M_{\rm NS}$) $1.4\,M_{\odot}$ and radius ($R_{\rm NS}$) 10\,km. We then obtain an upper limit for mass accretion rate of $3.96 \times 10^{-9}\,M_{\odot}\ y^{-1}$ for $D = 6.3\,\rm kpc$.


The reflection model shows that the inner edge of the accretion disc extends inwards to $R_{\rm in} = 5.41-6.40\,R_{\rm g}$ (11.20--13.24\,km), which is near the surface of the neutron star. If the inner disc is truncated at the magnetosphere radius, we used Equation (1) of \citet{2009Cackett} to calculate the upper limit of the magnetic field strength of the NS:
\begin{equation}
\begin{aligned}
\mu = & 3.5 \times 10^{23} k_{\rm A}^{-7 / 4} x^{7 / 4}\left(\frac{M}{1.4 }\right)^2 \\
& \times\left(\frac{f_{\rm a n g}}{\eta} \frac{F_{\rm b o l}}{10^{-9}\ {\rm ergs}\ {\rm cm}^{-2}\ \rm{s}^{-1}}\right)^{1 / 2} \frac{D}{3.5\ {\rm kpc}}\ {\rm G}\ {\rm cm}^3
\label{eq:flied}
\end{aligned}
\end{equation}
where $k_{\rm A}$ is the coefficient depending on the conversion from spherical to disk accretion, $f_{\rm ang}$ is the anisotropy correction factor \citep[]{2009Ibragimov}, and $\eta$ is the accretion efficiency in the Schwarzschild metric. The bolometric flux of $F_{\rm b o l} = 7.32 \times 10^{-9}\ \rm ergs\ \rm cm^{-2}\ \rm s^{-1}$  is estimated by extrapolating the spectral fit over the 0.1--100\,keV range. We set $k_{\rm A} = 0.5$, $f_{\rm ang} = 1$, and $\eta = 0.1$ for an NS of mass 1.4\,$M_{\rm \odot}$ and radius 10\,km based on \citet{2009Cackett}. $x$ is obtained from $R_{\rm in} = xGM/c^{2}$. If we take the constraint of $x = 6.4$ and $D = 6.3\,\rm kpc$, the magnetic field strength is $B \leq 4.67 \times 10^{8}\,\rm G$, which is similar with other AMXPs, and consistent with the calculation in \cite{2023Li} by assuming that the spin-up of the pulsar is solely induced by the angular momentum transferred from the accreted material to the NS.


\subsection{The \textasciitilde2.5\,Hz modulation generation mechanism}

As shown in Fig.~\ref{fig:PDS_3}, we find that MAXI J1816--195 has a broad peak at low frequencies in the PDS and does not find any kHz QPO. The PDS of normal atoll sources is a more complex at low frequencies (< 200\,Hz) characterized by broadband-limited noise and broad Lorentzian components \citep[]{2002Belloni}. In the case of MAXI J1816--195, the PDS shows a peak that, although not as sharp as the classic QPO, is concentrated in the 0.1--10\,Hz range, with Poisson noise dominating the rest of the frequency range. Similar peaks in the PDS have also been observed in SAX J1808.4--3658 and NGC 6440 X--2, and the Gaussian component with the highest peak power in the PDS has been identified as a QPO. Additionally, \cite{Di_Salvo_2001} found evidence that a component can sometimes switch from being a broad noise-like component into a QPO or vice versa. Therefore, we conclude that the strong \textasciitilde2.5\,Hz modulation observed in MAXI J1816--195 is most likely a low-frequency QPO rather than a broadband-limited noise.



However, the observational properties of this \textasciitilde2.5\,Hz modulation differ from those of the low-frequency QPOs observed in most AMXPs. The rms of MAXI J1816--195 increases with energy, but the QPOs of Aql X-1 and IGR J00291+5934 are displayed in soft X-rays with fractional rms falling strongly with energy. Another difference is the connection with the burst; the QPOs of Aql X--1 disappeared after the burst, suggesting that QPOs could be related to the stable burning in the neutron star envelope. The burst does not affect the \textasciitilde2.5\,Hz modulation of MAXI 1816--195. Thus, implying a different mechanism for this \textasciitilde2.5\,Hz modulation. The QPO amplitudes of MAXI J0911--655 increase with energy, but no QPOs can be resolved for the 10--30\,keV energy band \citep[]{2017BultB}. Instead, only a single broad noise term is present. Results of SAX J1808.4--3658 and NGC 6440 X--2 are most similar to the characteristics of MAXI J1816--195. The QPO frequencies are very close to the frequency of MAXI J1816--195 and the energy dependence of the QPO amplitude also increases with energy. \citet{patruno2009} and \citet{patruno2013} interpreted the modulation as hydrodynamic disk instabilities when the system moves into the propeller regime at very low $\dot{M}_{\rm \odot}$. In the propeller regime, the Keplerian velocity in the innermost region of the accretion disk is slower than the rotational velocity of the neutron star magnetosphere. While the innermost radius of the accretion disk of MAXI J1816--195 is very small, it does not meet the conditions of the propeller stage (the corotation radius is \textasciitilde 26\,km for MAXI J1816--195). Moreover, we do not observe flux drop in the light curve. So, the trapped disk instability model cannot explain the \textasciitilde2.5\,Hz modulation we found.

The fractional rms of the \textasciitilde2.5\,Hz modulation in MAXI J1816--195 increased with energy from \textasciitilde 12\% at 2\,keV to \textasciitilde 30\% at 8\,keV (see Fig.~\ref{fig:rms_energy_7}), while at higher energies it still remained at a high value. Additionally, given that there are insignificant oscillations below 2\,keV, and that the appearance of \textasciitilde2.5\,Hz modulation (see Fig.~\ref{fig:hid_2} and Fig.~\ref{fig:PDS_lc_5}) is strongly correlated with a harder spectrum, it is suggested that this \textasciitilde2.5\,Hz modulation may be produced by the hard component, perhaps from an unstable corona. From Fig.~\ref{fig:dps_burst_9}, it can be seen that the modulation is almost unaffected by the burst, which suggests that the location where the modulation is generated should be relatively far from the neutron star, tending towards a disk-corona. If assuming a lamppost geometry of the corona, via fitting the spectra with the {\tt relxilllpcp} model, the value of corona height ($h$) is several $R_{\rm g}$ (see Table~\ref{tab:app} in the appendix), providing further evidence for the disk-corona model. This is because it would be challenging for a lamppost corona to remain unaffected at such a low height during the burst. In addition, \cite{2022Chen} mentioned that MAXI J1816--195 was significantly cooled during the burst, by considering that the photons in the 30--100\,keV range during the burst peak accounted for only \textasciitilde 30\% of the persistent flux, which indicates that temperature change of the corona is not the main factor accounting for the \textasciitilde2.5\,Hz modulation generation. \cite{2022Bellavita} used a variable Comptonization model ({\tt vKompth}) to explore the behavior of the spectra by independently varying the corona parameters. From their Figure 3, it can also be seen that changes in the electron temperature have little effect on the amplitude of the \textasciitilde2.5\,Hz modulation above 2\,keV, while the photon index (with the electron temperature fixed), which represents the optical depth, is the main parameter affecting the amplitude of the \textasciitilde2.5\,Hz modulation. Therefore, the \textasciitilde2.5\,Hz modulation of MAXI J1816--195 is likely caused by oscillations in the optical depth of the corona, either caused by the variations of the plasma density or by the corona size, or both. Together with the slight increase in intensity for periods with the \textasciitilde 2.5\,Hz modulation (Fig.~\ref{fig:PDS_lc_5}) and the intermittent phenomenon of the \textasciitilde2.5\,Hz modulation, we suggest that the oscillations in the optical depth are further likely to be related to the accretion variability of the neutron star.

\section*{Acknowledgements}

We thank the anonymous referee for useful comments that have improved the paper. This work made use of the data and software from the High Energy Astrophysics Science Archive Research Center (HEASARC), provided by NASA’s Goddard Space Flight Center,and the \textit{Insight}-HXMT mission, funded by China National Space Administration (CNSA) and the Chinese Academy of Sciences (CAS). This work is supported by the National Key R\&D Program of China (2021YFA0718500). We acknowledge funding support from the National Natural Science Foundation of China (NSFC) under grant No. 12122306, No. U2038102, No. U2031205, No. U2038104, No. U1838201, No. U1838108, No. 12173103, the CAS Pioneer Hundred Talent Program Y8291130K2 and the Scientific and technological innovation project of IHEP Y7515570U1. Thanks to Thomas Dauser and Javier Garcia for their guidance on reflection models.

\section*{Data Availability}

The data used for this article are publicly available in the High Energy Astrophysics Science Archive Research Centre (HEASARC) at \url{https://heasarc.gsfc.nasa.gov/cgi-bin/W3Browse/w3browse.pl}



\bibliographystyle{mnras}
\bibliography{example} 




\FloatBarrier
\appendix
\section{Figures}
\label{A}
\begin{figure}
	\includegraphics[width=\columnwidth]{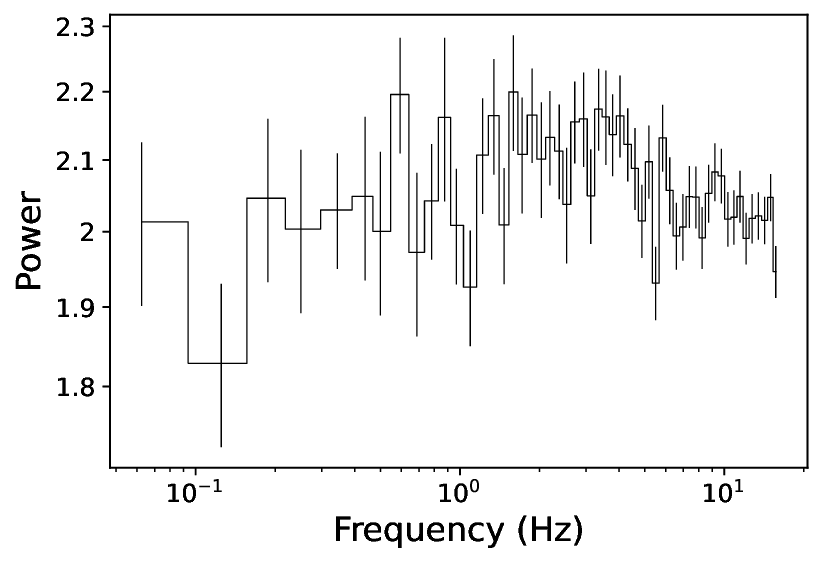}

    \caption{Leahy-normalized power density spectrum from \textit{NICER}’s ObsID 5533011501 in the 1.5--1.8\,keV band.}
  \label{fig:PDS1.5}
\end{figure}

\section{{\tt relxilllpCp} model fitting results}
\label{B}

We fit the broadband spectra of MAXI J1816--195 using {\tt relxilllpcp} with a lamp post geometry, and we fixed the energy of the two edges. The inclination angle is also fixed at 10. This does not affect the values of other spectral parameters. The best spectral parameter results using the {\tt relxilllpCp} model are presented in Table~\ref{tab:app} of the appendix.

\begin{table}

 \centering
 \caption{Best-fitting spectral parameters of the \textit{NICER} and \textit{NuSTAR} observations of MAXI J1816--195 using Model 3: {\tt const}$\times${\tt edge}$\times${\tt edge}$\times${\tt tbabs}$\times${\tt (relxilllpCp}$+${\tt diskbb)}. Uncertainties are at 90\%.}
 \renewcommand\arraystretch{1.36}
 \begin{tabular}{ccc}
   \hline
   Model & Parament (unit) & Value \\
   \hline
   TBABS & $N_{\rm H}\ (\times10^{22}\ {\rm cm^{-2}})$ & $2.28^{+0.02}_{-0.02}$ \\
   EDGE & $E_{\rm edge,1} $(keV)& $1.84$\ (fixed)\\
   EDGE & $E_{\rm edge,2} $ (keV) & $0.90$\ (fixed)\\
   DISKBB & $kT_{\rm in}\ ({\rm keV})$  & $0.558^{+0.002}_{-0.003}$  \\
      & $N_{\rm diskbb}$  &$870^{+37}_{-22}$ \\
   
   RELXILLLPCP & $\Gamma$ & $1.93^{+0.02}_{-0.01}$\\
    & $kT_{\rm e}\ ({\rm keV})$ & $27.84^{+0.64}_{-0.52}$\\
    & $R_{\rm in}$\ $(\times R_{\rm ISCO})$ & $1.11^{+0.04}_{-0.04}$\\
              & $h$\ ($R_{\rm g}$)&$< 3.2394$\\
    & $a$ & 0.248\ (fixed)\\
    & Refl\_frac & $0.21^{+0.01}_{-0.02}$  \\
    & $i\ (^{\circ})$ & $10$\ (fixed)  \\
    & $\log\xi$ & $3.01^{+0.09}_{-0.14}$ \\
    & $\log N\ ({\rm cm^{-3}})$ & $18.97^{+0.09}_{-0.36}$\\
    & $A_{\rm Fe}$\ $(\times \rm solar)$ & $2.52^{+0.18}_{-0.16}$ \\
    & Norm & $0.162^{+0.007}_{-0.003}$\\
   \hline
    & $\chi^2$/d.o.f & 1.03\ (1726/1680) \\
   \hline
 \end{tabular}
 \label{tab:app}
\end{table}


\bsp	
\label{lastpage}
\end{document}